\documentclass{aa}
\usepackage{graphicx}
\begin{document}

\authorrunning{K\"apyl\"a et al.}
\titlerunning{ Dynamo coefficients III }

   \title{Magnetoconvection and dynamo coefficients III:}

   \subtitle{$\alpha$-effect and magnetic pumping in the rapid rotation regime}

   \author{P. J. K\"apyl\"a
	  \inst{1}$^{,}$\inst{2},
	  M. J. Korpi
	  \inst{3},
	  M. Ossendrijver
	  \inst{2},
          \and
	  M. Stix
	  \inst{2}
	  }

   \offprints{P. J. K\"apyl\"a\\
	  \email{petri.kapyla@oulu.fi}
	  }

   \institute{Astronomy Division, Department of Physical Sciences,
              PO BOX 3000, FI-90014 University of
              Oulu, Finland
	  \and Kiepenheuer--Institut f\"ur Sonnenphysik, 
	      Sch\"oneckstrasse 6, D--79104 Freiburg, Germany
          \and Observatory, PO BOX 14, FI-00014 University of Helsinki, 
              Finland \\ }

   \date{Received 6 February 2006 / Accepted 2 May 2006}

\abstract{}
         {The $\alpha$- and $\gamma$-effects, which are responsible
           for the generation and turbulent pumping of large scale
           magnetic fields, respectively, due to passive advection by
           convection are determined in the rapid rotation regime
           corresponding to the deep layers of the solar convection
           zone.}
         {A 3D rectangular local model is used for solving the full
           set of MHD equations in order to compute the electromotive
           force (emf), $\vec{\mathcal{E}} = \overline{\vec{u} \times
             \vec{b}}$, generated by the interaction of imposed weak
           gradient-free magnetic fields and turbulent convection with
           varying rotational influence and latitude. By expanding the
           emf in terms of the mean magnetic field, $\mathcal{E}_i =
           a_{ij} \overline{B}_j$, all nine components of $a_{ij}$ are
           computed. The diagonal elements of $a_{ij}$ describe the
           $\alpha$-effect, whereas the off-diagonals represent
           magnetic pumping. The latter is essentially the advection
           of magnetic fields by means other than the underlying
           large-scale velocity field. Comparisons are made to
           analytical expressions of the coefficients derived under
           the first-order smoothing approximation (FOSA).}
         {In the rapid rotation regime the latitudinal dependence of
           the $\alpha$-components responsible for the generation of
           the azimuthal and radial fields does not exhibit a peak at
           the poles, as is the case for slow rotation, but at a
           latitude of about 30$\degr$. The magnetic pumping is
           predominantly radially down- and latitudinally equatorward
           as in earlier studies. The numerical results compare
           surprisingly well with analytical expressions derived under
           first-order smoothing, although the present calculations
           are expected to lie near the limits of the validity range
           of FOSA.}
         {}


   \maketitle


\section{Introduction}

   According to dynamo theory small-scale helical turbulence caused by
   the convective instability in combination with global rotation,
   uniform or differential, is the source of the large-scale magnetic
   structures seen in the Sun and in various other late-type stars
   (e.g. Parker \cite{Parker1955}; Moffatt \cite{Moffatt1978}; Krause
   \& R\"adler \cite{KrauRaed1980}). The small scales enter the
   evolution equation of the mean magnetic field via the turbulent
   electromotive force $\vec{\mathcal{E}} = \overline{\vec{u} \times
     \vec{b}}$. If the mean magnetic field $\overline{\vec{B}}$ varies
   slowly in time and space, the emf can be represented in terms of
   $\overline{\vec{B}}$ and its gradients
   \begin{equation}
     \mathcal{E}_i = a_{ij} \overline{B}_j + b_{ijk} \frac{\partial \overline{B}_j}{\partial x_k} + \ldots \;, \label{equ:mfemf}
   \end{equation}
   where $a_{ij}$ and $b_{ijk}$ are in the general case tensors
   containing the transport coefficients, and the dots indicate that
   higher order derivatives can be taken into account. The challenge
   is to derive the tensors $a_{ij}$ and $b_{ijk}$ which, in general,
   cannot be done from first principles due to the lack of a
   comprehensive theory of convective turbulence.

   In the kinematic regime where the magnetic energy is negligible in
   comparison to the kinetic energy the most simple approximation is
   to neglect all correlations higher than second order in the
   fluctuations. This is often called the first order smoothing
   (FOSA), quasilinear, or the second order correlation approximation
   (SOCA). Considering the simple case of isotropic turbulence in the
   high-conductivity limit as an example where the transport
   coeffients can be computed analytically, the tensor $a_{ij}$
   reduces into a single scalar (Steenbeck \& Krause
   \cite{SteenKrau1969})
   \begin{equation}
     \alpha = - \frac{1}{3} \tau_{\rm c} \overline{\vec{\omega} \cdot
       \vec{u}}\;, \label{equ:aisofosa}
   \end{equation}
   where $\tau_{\rm c}$ is the correlation time of the turbulence,
   $\vec{\omega} = \nabla \times \vec{u}$ the vorticity, and
   $\vec{\omega} \cdot \vec{u}$ the kinetic helicity. We note that in
   the anisotropic case the trace of the tensor $a_{ij}$, instead of
   any individual component of the $\alpha$-effect, is proportional to
   the kinetic helicity (e.g. R\"adler \cite{Raedler1980}; see also
   Sect.~\ref{subsec:comFOSA}).

   Despite the limited applicability of Eq.~(\ref{equ:aisofosa}),
   numerical models of magnetoconvection have shown that there is at
   least a qualitative agreement between the $\alpha$-effect and the
   negative of the kinetic helicity also in the anisotropic case
   (Brandenburg et al. \cite{Brandea1990}; Ossendrijver et
   al. \cite{Osseea2001}, hereafter Paper I). Furthermore, using local
   time-distance helioseismology, Gizon \& Duvall (\cite{GizDuv2003})
   studied the correlation of the curl and divergence of the
   horizontal components of solar surface flows as a proxy for the
   helicity (see also R\"udiger et al. \cite{Ruedigerea1999}) and
   found that the correlation follows a $\cos \theta$ latitude profile
   for the observed latitude range. Using local numerical convection
   models Egorov et al. (\cite{Egorovea2004}) were able to reproduce
   the $\cos \theta$ latitude profile and further confirmed the
   validity of the correlation as a valid tracer of the helicity. In a
   recent study, K\"apyl\"a et al. (\cite{Kaepylaeea2004}) found a
   similar latitude dependence for moderate rotation, i.e. up to
   Coriolis number ${\rm Co} = 2\,\Omega l/u < 4$, in agreement with
   the results of Egorov et al. For more rapid rotation, Co $\approx
   10$, however, the same authors found that the latitude distribution
   of the volume average over the convectively unstable region
   resembles more a $\sin \theta$ profile for latitudes higher than
   15$\degr$. As our working hypothesis, we consider the relation
   (\ref{equ:aisofosa}) to trace the $\alpha$-effect, which raises the
   question of the latitude dependence of this effect in the rapid
   rotation regime.

   In the present study we examine the $\alpha$-effect by means of
   local numerical modelling of small volumes of a star at different
   latitudes. In distinction to other studies (e.g. Brandenburg et
   al. \cite{Brandea1990}; Ossendrijver et al. \cite{Osseea2001},
   \cite{Osseea2002}, hereafter Paper II), we extend the parameter
   range in the rapid rotation regime to values typical for the base
   of the solar convection zone, where the Coriolis number is of the
   order of 10 or larger (see e.g. K\"uker et al. \cite{Kuekerea1993};
   K\"apyl\"a et al. \cite{Kaepylaeea2005b}).

   As a secondary objective we make an attempt to test the validity of
   FOSA. A sufficient, but not a necessary, requirement for the
   applicability of FOSA is that either the Reynolds or the Strouhal
   number is small
   \begin{equation}
     {\rm min} \bigg(\frac{ul}{\eta}, u k_{\rm f} \tau_{\rm c} \bigg)
     = {\rm min (Rm, St)} \ll 1\;, \label{equ:fosarequ}
   \end{equation}
   where $u$ and $l$ are the typical velocity and length scales,
   $k_{\rm f}$ the wavenumber of the energy carrying scale which is
   related with the correlation lenght via $l_{\rm c} = 2\pi/k_{\rm
     f}$. In the high-conductivity limit relevant for
   Eq.~(\ref{equ:aisofosa}), where $\eta \tau_{\rm c}/l_{\rm c}^2 \ll
   1$, the requirement (\ref{equ:fosarequ}) reduces simply to ${\rm
     St} \ll 1$ which, however, is at best marginally satisfied
   (e.g. Brandenburg et al. \cite{Brandea2004}; Brandenburg \&
   Subramanian \cite{BranSub2005b}; K\"apyl\"a et
   al. \cite{Kaepylaeea2006}).

   We derive analytical formulae for all of the $a_{ij}$-coefficients
   in the high-conductivity limit. For the different components of the
   $\alpha$-effect we find expressions analogous to
   Eq.~(\ref{equ:aisofosa}) in the anisotropic case. We then compare
   the analytical result and its numerical equivalent using the
   correlation time as a free parameter. The correlation time allows
   us, at least in principle, to compute the Strouhal number.

   The remainder of the paper is organised as follows: in
   Sect.~\ref{sec:model} the numerical convection model is
   described. In Sect.~\ref{sec:alphares} the problem and research
   methods are stated. In Sects.~\ref{sec:results} and
   \ref{sec:conclu} we give the results and conclusions, respectively.


\section{The convection model}
\label{sec:model}

\subsection{Basic equations}
   The convection model is the same as that presented in K\"apyl\"a et
   al. (\cite{Kaepylaeea2004}, \cite{Kaepylaeea2005b}). The
   computational domain is a rectangular box at latitude $\Theta$ in
   the southern hemisphere of a star. The coordinate system of the
   model is chosen such that $x$, $y$, and $z$ correspond to
   $(-\theta,\phi,-r)$ in spherical coordinates, where $\theta =
   90\degr - \Theta$ is the colatitude. The angular velocity as a
   function of latitude is given by $\vec{\Omega} = \Omega(\cos \Theta
   \vec{\hat{{e}}_{x}} - \sin \Theta \vec{\hat{{e}}_{z}})$. The box
   has horizontal dimensions $L_{\rm x} = L_{\rm y} = 4$, and $L_{\rm
     z} = 2$ in the vertical direction in units of the depth of the
   convectively unstable layer $d$. We set $(z_0, z_1, z_2, z_3) =
   (-0.15,0,1,1.85)$, where $z_0$ and $z_3$ correspond to the top and
   bottom boundaries of the box, respectively, whereas $z_1$ and $z_2$
   give the positions of the upper and lower boundaries of the
   convectively unstable layer. We solve a system of MHD-equations
   \begin{eqnarray}
   \frac{\partial {\vec A}}{\partial t} &=& {\vec u} \times {\vec B} - \eta \mu_{0}{\vec J}\;, \\
   \frac{\partial \ln \rho}{\partial t} &=& - ({\vec u} \cdot \nabla)\ln \rho + \nabla \cdot {\vec u}\;, \\
   \frac{\partial {\vec u}}{\partial t} &=& - ({\vec u} \cdot \nabla){\vec u} - \frac{1}{\rho}\nabla p - 2\,\vec{\Omega} \times {\vec u} + {\vec g} + \nonumber \\
   && \hspace{2.5cm} + \frac{1}{\rho}{\vec J} \times {\vec B} + \frac{1}{\rho} \nabla \cdot (2 \nu \rho \tens{S})\;,\label{equ:momentum}\\
   \frac{\partial e}{\partial t}  &=&  - ({\vec u} \cdot \nabla)e - \frac{p}{\rho}(\nabla \cdot {\vec u}) + \frac{\eta \mu_0}{\rho} \vec{J}^2 + \nonumber \label{equ:ee} \\
   && \hspace{2.5cm} + \Gamma_{\rm cond} + \Gamma_{\rm visc} - \Gamma_{\rm cool}\;,
   \end{eqnarray}
   where $\vec{A}$ is the magnetic vector potential, $\vec{u}$ the
   velocity, $\vec{B} = \nabla \times \vec{A} + \vec{B}_0$ the
   magnetic field which is the sum of the fluctuating and (constant)
   imposed contributions, $\vec{J} = \nabla \times \vec{B}/\mu_0$ the
   current density, $\rho$ the mass density, $p$ the pressure,
   $\vec{g} = g\vec{\hat{e}}_z$ the constant gravity,
   $\vec{\hat{e}}_z$ the unit vector in the vertical direction,
   $\tens{S}_{ij} = \frac{1}{2}(u_{i,j} + u_{j,i} -
   \frac{2}{3}\delta_{ij} \nabla \cdot \vec{u})$ the strain tensor,
   and $e = c_{\rm V}T$ the internal energy per unit mass. $\mu_0$ is
   the vacuum permeability, $\eta$ and $\nu$ the constant magnetic
   diffusivity and kinematic viscosity. $\Gamma_{\rm cond}$ describes
   thermal conduction (see the next subsection). The term $\Gamma_{\rm
     visc}$ describes viscous heating via
   \begin{eqnarray}
   \Gamma_{\rm visc} = 2 \nu \tens{S}_{ij} \frac{\partial
     u_i}{\partial x_j}\;.
   \end{eqnarray}
   The uppermost layer of the box is cooled according to
   \begin{eqnarray}
   \Gamma_{\rm cool} = \frac{1}{t_{\rm cool}} f(z) (e - e_{0})\;,
   \label{equ:cool}
   \end{eqnarray}
   where $t_{\rm cool}$ is a cooling time, chosen to be so short that
   the upper boundary stays isothermal, $f(z) = (z - z_1)/(z_0 - z_1)$
   a function which is applied in the interval $z_{0} \le z < z_{1}$,
   and $e_{0} = e(z_{0})$ the value of internal energy (temperature)
   at the top of the box. The cooling term serves as a parametrisation
   of the radiative losses occuring at the surface of the star. We
   assume an ideal gas with
   \begin{eqnarray}
   p = \rho e (\gamma - 1)\;,
   \end{eqnarray}
   where the ratio of the specific heats $\gamma = c_{\rm P}/c_{\rm V}
   = 5/3$. $e_0$ and the stratification in the box are fixed by the
   parameter
   \begin{eqnarray}
     \xi_0 = \frac{(\gamma-1) e_0}{gd}\;, 
   \end{eqnarray}
   which defines the pressure scale height at the top of the box.

\subsection{Boundary conditions and initial setup}
   We adopt periodic boundary conditions in the horizontal directions,
   and closed stress free boundaries at the top and at the bottom
   \begin{eqnarray}
     \frac{\partial u_{x}}{\partial z} = \frac{\partial u_{y}}{\partial z} = u_{z} &=& 0\; \hspace{2cm} {\rm at} \;\;\;z = z_{0},z_{3}\;.
   \end{eqnarray}
   For the vector potential we use
   \begin{eqnarray}
     \frac{\partial A_{\rm x}}{\partial z} = \frac{\partial A_{\rm y}}{\partial z} = A_z = 0\; \hspace{2cm} {\rm at} \;\;\;z = z_{0},z_{3}\;,
   \end{eqnarray}
   which constrain the fluctuating magnetic field to be vertical at
   the boundaries. We split the heat conduction term into two parts
   \begin{eqnarray}
     \Gamma_{\rm cond} = \nabla \cdot [\kappa_{\rm t} \nabla (e - \overline{e}) + \kappa_{\rm h} \nabla \overline{e}]\;,
   \end{eqnarray}
   where the first term acts only on the fluctuations and the latter
   only on the mean, i.e. horizontally averaged, stratification. Thus
   $\kappa_{\rm t}$ and $\kappa_{\rm h}$ can be considered as the
   turbulent and radiative conductivities, which satisfy $\kappa_{\rm
     t} \gg \kappa_{\rm h}$ in real stars. We define the conductivies
   as
   \begin{eqnarray}
     \kappa_{\rm t} &=& \gamma \rho \chi_0\;, \\
     \kappa_{\rm h} &=& \frac{(\gamma-1) F_{\rm r}}{g \nabla} \;, \label{equ:kappah}
   \end{eqnarray}
   where $\chi_0$ is the reference value of the thermal diffusivity,
   $F_{\rm r}$ the input energy flux, and $\nabla$ the mean
   logarithmic temperature gradient in the initial state.

   As a boundary condition we fix the temperature at the top of the
   box and apply a constant mean temperature gradient at the bottom
   \begin{eqnarray}
     e_{z = z_{0}} &=& e_{0}\;, \\
     \frac{\partial e}{\partial z}_{z = z_{3}} &=& \frac{g}{(\gamma - 1)} \nabla_3\;,
   \label{equ:dez}
   \end{eqnarray}
   where $\nabla_3$ is the logarithmic temperature gradient at the
   lower boundary (see below).

   Initially the radiative flux, $\vec{F}_{\rm rad} = \kappa_{\rm h}
   \nabla \overline{e}$, carries the whole energy through the
   domain. In the present paper the uppermost layer is initally
   isothermal, and the stratification in the convectively unstable
   region is described by a polytropic index $m_2$. Often $m_2 = 1$ is
   chosen, but this has the disadvantage that the time needed to reach
   the final thermally relaxed state which is much closer to the
   adiabatic stratification becomes large (of the order of $t_{\rm
     relax} = \rho \gamma d^2/\kappa_{h} \approx 4 \cdot 10^3
   \sqrt{d/g}$ in the present study, see also Brandenburg et
   al. \cite{Brandea2005}). Thus in practice, only the conductivity
   $\kappa_{\rm h}$ strictly corresponds to the case $m_2 = 1$ in the
   initial state, but the thermal stratification corresponds to
   polytropic index $m_2 = 1.25$ which is already close to the
   thermally saturated state. The initial temperature gradient in the
   lower part of the convectively unstable layer and the overshoot
   layer is calculated from
   \begin{eqnarray}
     \nabla(z) = \nabla_{\rm 3} + \frac{1}{2} \{\tanh [4(z_{\rm m} - z)] + 1\}\Delta \nabla\;,
   \end{eqnarray}
   where $\nabla_3$ is the imposed gradient, and $\Delta \nabla =
   \nabla_2 - \nabla_3$ the difference between the temperature
   gradients in the convectively unstable layer and the bottom
   boundary. $z_{\rm m}$ is calculated on the condition that $\nabla =
   \nabla_{\rm ad} = (\gamma-1)/\gamma$ at $z = z_2$ in the initial
   non-convecting state. The density stratification is obtained via
   the equation of hydrostatic equilibrium.

   Furthermore, we decrease the input energy flux by a factor of
   roughly ten in comparison to Papers I and II. Decreasing $F_{\rm
     r}$ even further would be desirable, given the fact that the
   ratio of the input flux in the model to the solar energy flux is
   still $F_{\rm r}/F_\odot \approx 10^{7}$. However, from
   Eq.~(\ref{equ:kappah}) it is seen that $F_{\rm r} \propto \kappa_h$
   which leads to $t_{\rm relax} \propto F_{\rm r}^{-1}$. Thus $F_{\rm
     r}$ cannot be decreased by a very large amount without a
   prohibitively large increase in the needed computational time. We
   consider the present setup to be a good compromise between real
   stellar conditions and the computational resources required.

\subsection{Dimensionless quantities}

   We measure length with respect to the depth of the unstable layer,
   $d = z_{2} - z_{1}$, density in units of the initital value at the
   bottom of the convectively unstable layer, $\rho_{0}$, and
   acceleration in units of the gravitational acceleration
   $g$. Futhermore, magnetic permeability is measured in units of
   $\mu_0$ and entropy in terms of $c_{\rm P}$. From these choices it
   follows that time is measured in units of the free fall time,
   $\sqrt{d/g}$, velocity in units of $\sqrt{dg}$, magnetic field in
   terms of $\sqrt{d\rho_{0}\mu_{0}g}$, and temperature in terms of
   $gd/c_{\rm P}$.

   The calculations are controlled by the following dimensionless
   parameters. The relative strengths of the thermal and magnetic
   diffusion against kinetic diffusion are measured by the kinetic and
   magnetic Prandtl numbers
   \begin{eqnarray}
     {\rm Pr} &=& \frac{\chi_{0}}{\nu}\;, \label{equ:Pr}\\
     {\rm Pm} &=& \frac{\eta}{\nu}\;,
   \end{eqnarray}
   where $\chi_{0}$ is the reference value of the thermal diffusivity
   taken from the middle of the unstably stratified layer.

   Rotation is measured by the Taylor number
   \begin{eqnarray}
     {\rm Ta} = \Big( \frac{2\,\Omega d^{2}}{\nu} \Big)^{2}\;.
   \end{eqnarray}
   A related dimensionless quantity is the Coriolis number, which is
   the inverse of the Rossby number, Co = $2\,\Omega \tau$, where
   $\tau = d/u_{\rm rms}$ is an estimate of the convective turnover
   time realised in the calculation.

   Convection efficiency is measured by the Rayleigh number
   \begin{eqnarray}
     {\rm Ra} = \frac{d^{4}g\delta}{\chi_{0}\nu H_{\rm ph}}\;,
   \end{eqnarray}
   where $\delta = \nabla - \nabla_{a}$ is the superadiabaticity,
   measured as the difference between the radiative and the adiabatic
   logarithmic temperature gradients, and $H_{\rm ph}$ is the pressure
   scale height, both evaluated from the non-convecting initial state
   in the middle of the unstably stratified layer.

   The strength of the imposed magnetic field is described by the
   Chandrasekhar number
   \begin{eqnarray}
     {\rm Ch} = \frac{\mu_{0} B_{0}^{2} d^{2}}{4 \pi \rho_{0} \nu
       \eta}\;, \label{equ:Ch}
   \end{eqnarray}
   where $B_0$ is the magnitude of the imposed field.

   We measure the strengths of the advection terms versus the
   corresponding diffusion terms in the momentum and induction
   equations by the Reynolds numbers
   \begin{eqnarray}
     {\rm Re} = \frac{u_{\rm rms} d}{\nu}\;,\\
     {\rm Rm} = \frac{u_{\rm rms} d}{\eta}\;,
   \end{eqnarray}
   where $u_{\rm rms}$ is the rms value of the velocity fluctuations
   in the convectively unstable layer. The parameters Pr, Pm, Ta, Ra,
   and Ch are used as inputs to the model, whereas Re, Rm, and Co are
   values computed from the actual runs.

   The numerical method used is a modified version of that presented
   in Caunt \& Korpi (\cite{CauKo2001}), employing sixth-order
   accurate explicit finite differences and a third-order accurate
   Adams-Bashforth-Moulton time stepping routine. The code is
   parallelised using message passing interface (MPI). The
   calculations were carried out on the 34-processor BAGDAD Linux
   cluster hosted by the Kiepenheuer-Institut f\"ur Sonnenphysik.

   \begin{table}
   \centering
      \caption[]{Summary of the calculations. The following parameters
        are common for all runs: $\nu = \eta = 2 \cdot 10^{-4}
        \sqrt{gd^3}$, $F_{\rm r} = 3 \cdot 10^{-4} \rho_0 (gd)^{3/2}$,
        $\nabla_3 = 0.375$, $\xi_0 = 0.2$, $B_0 = 10^{-4} \sqrt{d
          \rho_0 \mu_0 g}$, ${\rm Pr} = 0.4$, ${\rm Pm} = 1$, $\delta
        \approx 0.0432$, and $H_{\rm ph} \approx 0.422d$. These
        choices result in ${\rm Ch} = 0.02$ and ${\rm Ra} \approx 1.03
        \cdot 10^6$. A grid of $128^3$ is used. The quantities Re, Rm,
        Co and $u_{\rm rms}$ are averages over the convective layer
        and time, whereas Ta and $\Theta$ are fixed. We estimate that
        $\eta \tau_{\rm c}/l_{\rm c}^2 \approx 10^{-3}$ in all of the
        calculations, see Sect.~\ref{subsec:comFOSA}.}
      \vspace{-0.5cm}
      \label{tab:Calcu}
     $$
         \begin{array}{p{0.15\linewidth}ccccc}
            \hline
            \noalign{\smallskip}
            Run      & {\rm Re, Rm} & {\rm Ta} & \hspace{0.35cm}{\rm Co}\hspace{0.35cm} & u_{\rm rms} & \Theta \\
            \noalign{\smallskip}
            \hline
            \noalign{\smallskip}
	    Co1-90  & 257 & 6.5 \cdot 10^{4} & 1.02 & 0.051 & -90\degr \\
	    Co1-60  & 260 & 6.5 \cdot 10^{4} & 1.00 & 0.052 & -60\degr \\
	    Co1-30  & 264 & 6.5 \cdot 10^{4} & 0.99 & 0.053 & -30\degr \\
	    Co1-00  & 253 & 6.5 \cdot 10^{4} & 1.02 & 0.051 & \,\,\,0\degr \\
	    \hline
            \noalign{\smallskip}
	    Co4-90  & 241 & 1.0 \cdot 10^{6} & 4.40 & 0.048 & -90\degr \\
	    Co4-60  & 247 & 1.0 \cdot 10^{6} & 4.38 & 0.049 & -60\degr \\
	    Co4-30  & 267 & 1.0 \cdot 10^{6} & 4.00 & 0.054 & -30\degr \\
	    Co4-00  & 319 & 1.0 \cdot 10^{6} & 3.30 & 0.064 & \,\,\,0\degr \\
	    \hline
            \noalign{\smallskip}
	    Co10-90 & 209 & 6.5 \cdot 10^{6} & 13.2 & 0.042 & -90\degr \\
	    Co10-75 & 211 & 6.5 \cdot 10^{6} & 13.0 & 0.042 & -75\degr \\
	    Co10-60 & 213 & 6.5 \cdot 10^{6} & 12.8 & 0.043 & -60\degr \\
	    Co10-45 & 222 & 6.5 \cdot 10^{6} & 12.3 & 0.044 & -45\degr \\
	    Co10-30 & 250 & 6.5 \cdot 10^{6} & 10.7 & 0.050 & -30\degr \\
	    Co10-15 & 283 & 6.5 \cdot 10^{6} & 9.38 & 0.057 & -15\degr \\
	    Co10-00 & 408 & 6.5 \cdot 10^{6} & 7.23 & 0.082 & \,\,\,0\degr \\
            \noalign{\smallskip}
            \hline
         \end{array}
     $$ 
   \end{table}


   \section{The $a_{ij}$-tensor and methods of analysis}
   \label{sec:alphares}

   \subsection{The $a_{ij}$-tensor}
   In the present study we concentrate only on the first term of the
   expansion (Eq.~\ref{equ:mfemf})
   \begin{eqnarray}
     \mathcal{E}_i = a_{ij} \overline{B}_j\;,
   \end{eqnarray}
   which, in the present Cartesian geometry, can be written in matrix
   form
   \begin{eqnarray}
     \vec{\mathcal{E}} & = &  \left( \begin{array}{ccc}
a_{xx} & a_{xy} & a_{xz} \\
a_{yx} & a_{yy} & a_{yz} \\
a_{zx} & a_{zy} & a_{zz} \\
\end{array} \right)
     \left( \begin{array}{c}
\overline{B}_x(z) \\
\overline{B}_y(z) \\
\overline{B}_z \\
\end{array} \right)\;,
   \end{eqnarray}
   where we note that the horizontal mean magnetic fields can vary as
   function of depth due to the contributions generated by the fluid
   motions in the course of the calculation. It suffices to make three
   calculations with uniform magnetic fields imposed along each
   coordinate axis to obtain all nine components of the
   $a_{ij}$-tensor. This is done separately for each rotation rate and
   latitude.

   Separating the symmetric and antisymmetric parts of the tensor one
   arrives at (see e.g. R\"adler \cite{Raedler1980})
   \begin{eqnarray}
     \alpha_{ij} & = & (a_{ij} + a_{ji})/2\;, \\
     \gamma_{i} & = & -\varepsilon_{ijk}a_{jk}/2\;, \label{equ:gpump}
   \end{eqnarray}
   where $\alpha_{ij}$ is a symmetric tensor whose diagonal components
   describe the generation of the mean magnetic field and the
   off-diagonals contribute to the pumping effect. $\gamma_{i}$
   encompasses the antisymmetric contribution and describes the part
   of the pumping effect analogous to the advection term in the
   induction equation. More explicitly, the electromotive force can
   now be written as
   \begin{eqnarray}
     \vec{\mathcal{E}} = \vec{\alpha} \overline{\vec{B}} + \vec{\gamma} \times \overline{\vec{B}}\;, \label{equ:emfnondiff}
   \end{eqnarray}
   where $\vec{\alpha}$ is a symmetric tensor of second rank and
   $\vec{\gamma}$ the vector defined in Eq.~(\ref{equ:gpump}). The
   impact of the off-diagonal components of $\vec{\alpha}$ on the
   pumping effect can most easily be seen by writing (Paper II)
   \begin{eqnarray}
     \vec{\mathcal{E}} = \vec{\alpha}^D \cdot \overline{\vec{B}} + \vec{\gamma}^{(x)} \times \overline{\vec{B}}_x + \vec{\gamma}^{(y)} \times \overline{\vec{B}}_y + \vec{\gamma}^{(z)} \times \overline{\vec{B}}_z \label{equ:emfnondiffcar}\;,
   \end{eqnarray}
   where $\overline{\vec{B}}_i = \vec{\hat{e}}_i \overline{B}_i$, and
   \begin{eqnarray}
     \vec{\alpha}^D & = & \left( \begin{array}{c}
\alpha_{xx} \\
\alpha_{yy} \\
\alpha_{zz} \label{equ:alphaD} \\
\end{array} \right)\;,\\
     \vec{\gamma}^{(x)} & = & \vec{\gamma} + \left( \begin{array}{c}
0 \\
-\alpha_{xz} \\
\alpha_{xy} \\
\end{array} \right)\;, \label{equ:gammax}\\
     \vec{\gamma}^{(y)} & = & \vec{\gamma} + \left( \begin{array}{c}
\alpha_{yz}\\
0 \\
-\alpha_{xy} \\
\end{array} \right)\;,\label{equ:gammay}\\
     \vec{\gamma}^{(z)} & = & \vec{\gamma} + \left( \begin{array}{c}
-\alpha_{yz} \\
\alpha_{xz} \\
0 \\
\end{array} \right)\;.\label{equ:gammaz}
   \end{eqnarray}
   Eqs.~(\ref{equ:gammax}) to (\ref{equ:gammaz}) show that the pumping
   velocity can be different for different magnetic field components
   (Kichatinov \cite{Kichatinov1991}, Paper II). Note that in contrast
   to Eq.~(\ref{equ:emfnondiff}), which is independent of the chosen
   coordinate system, the coefficients $\vec{\alpha}^D$ and
   $\vec{\gamma}^{(i)}$ in the coordinate dependent
   Eq.~(\ref{equ:emfnondiffcar}) are not vectors on account of their
   transformation properties.

   In what follows, we refer to the $\alpha$-effect as the diagonal
   part of $a_{ij}$ defined through Eq.~(\ref{equ:alphaD}), in
   distinction to many earlier studies where it is considered as the
   full $\vec{\alpha} \overline{\vec{B}}$ contribution to
   Eq.~(\ref{equ:emfnondiff}) (e.g. Krause \& R\"adler
   \cite{KrauRaed1980}; R\"adler \cite{Raedler1980}; R\"adler \&
   Stepanov \cite{RaedlerStepanov2006a}). Furthermore, $\vec{\gamma}$,
   defined as the off-diagonal part of $a_{ij}$ via
   Eq.~(\ref{equ:gpump}), is from now on referred to as the general
   pumping effect, whereas the $\vec{\gamma}^{(i)}$, defined via
   Eqs.~(\ref{equ:gammax})-(\ref{equ:gammaz}), describe the
   field-direction dependent pumping effect.

   \begin{figure}
   \centering
   \includegraphics[width=0.5\textwidth]{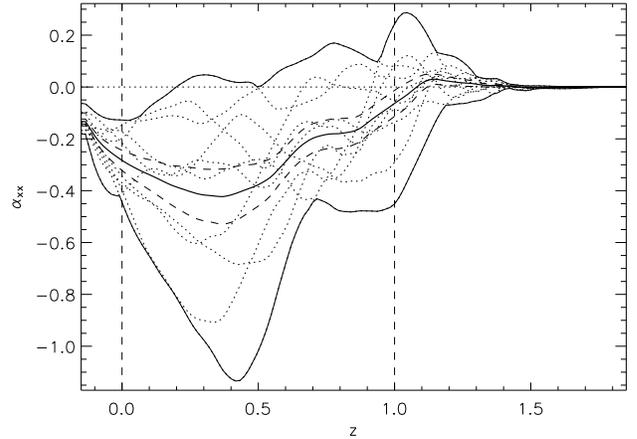}
      \caption{Variation of $\alpha_{xx} \approx
        \mathcal{E}_x/\overline{B}_x(z)$ in units of $0.01 \sqrt{dg}$
        in the run Co1-90 with an imposed field in the
        $x$-direction. Note that $\overline{B}_z(z) = 0$,
        max($|\overline{B}_y(z)|/|\overline{B}_x(z)|$) =
        $\mathcal{O}(0.1)$, and max($|\frac{ \partial
          \overline{B}_x(z)}{\partial z}|/|\overline{B}_x(z)|$) =
        $\mathcal{O}(0.1d^{-1})$, indicating that their influence on
        $\alpha_{xx}$ is only minor. The dotted curves show the
        average values from each of the 10 individual
        subintervals. The thick solid curve shows the average of the
        10 curves, and the dashed lines indicate the error estimates
        computed from the standard deviation $\sigma$ of the
        subaverages according to $\sigma/\sqrt{N}$. The thin solid
        lines indicate the full range, denoted from now on by
        $\tilde{\sigma}$, of values attained in the $N = 10$
        subintervals. The vertical dashed lines at $z = 0$ and $z = 1$
        denote the top and bottom of the convectively unstable region,
        respectively.}
      \label{pic:variation}
   \end{figure}

   \begin{figure*}
   \centering
   \includegraphics[width=1.\textwidth]{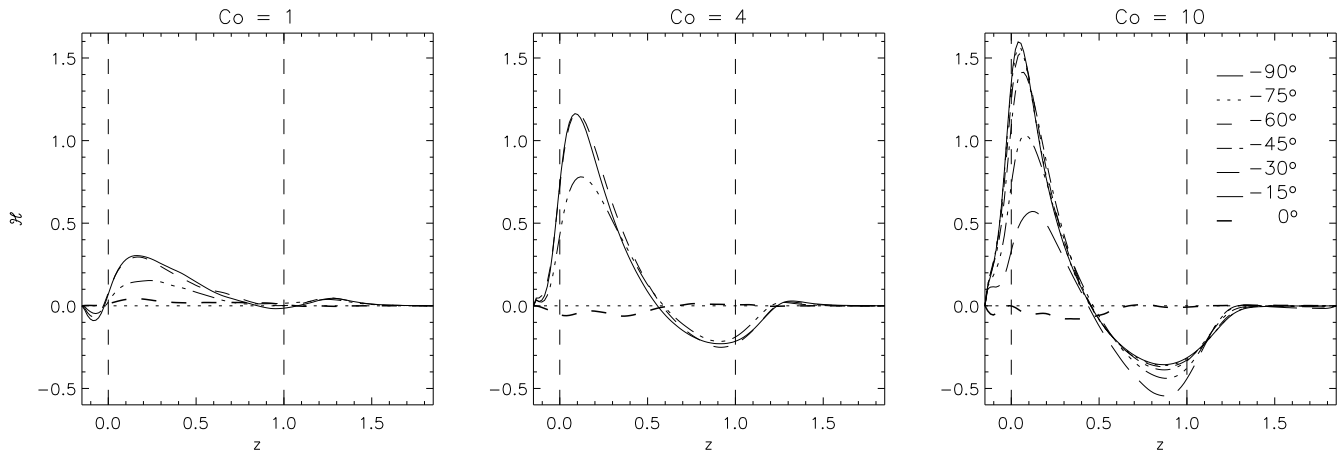}
      \caption{Kinetic helicity $\mathcal{H} = \overline{\vec{\omega}
          \cdot \vec{u}}$ in units of $0.01\,g$ for the approximate
        Coriolis numbers 1 (left), 4 (middle), and 10 (right). Each
        curve corresponds to a different latitude as indicated in the
        rightmost panel. The top and bottom of the convectively
        unstable regions are denoted by dashed vertical lines.}
      \label{pic:kinhel}
   \end{figure*}

   \subsection{Summary of the computations and averaging techniques}
   Table \ref{tab:Calcu} summarises the calculations. A parent run was
   evolved without rotation or magnetic fields until convection
   achieved a statistically steady state. In the present case the
   length of the parent run is $t = 100$ in units of $\sqrt{d/g}$. The
   final state of this run was used as an initial condition for the
   runs in Table \ref{tab:Calcu} for which an appropriate rotation
   vector was imposed. These purely hydrodynamical runs were evolved
   up to $t_0 = 500 \sqrt{d/g}$ in order to minimize the effects of
   the transients due to the added rotation and to allow time for
   thermal relaxation. After this an uniform magnetic field was added.

   Similarly as in Paper II we reset the fluctuating component of the
   magnetic field after every $T = 50 \sqrt{d/g}$ time units in order
   to minimize the contamination caused by the generated gradients of
   mean fields. The resetting interval is two times longer than in
   Paper II, which is justified by the fact that convection, i.e. the
   average rms velocities, in the present calculations are slower by
   approximately the same factor (see Table \ref{tab:Calcu}) due to
   the smaller input energy flux $F_{\rm r}$. In the averaging, the
   first $\Delta t_{\rm f} = 10 \sqrt{d/g}$ and the last $\Delta
   t_{\rm l} = 30 \sqrt{d/g}$ time units of each subinterval are
   neglected in accordance with the procedure of Paper II. The
   generated mean fields in this subinterval are also taken into
   account when computing the coefficients $a_{ij}$. As demonstrated
   in Paper II, only a few subintervals are needed for convergence. In
   the present study the results are averages over 10 subintervals.

   The horizontal and time average of a quantity $f$ over the ith
   subinterval is defined via
   \begin{eqnarray}
     &\overline{f}_i(z)& = \nonumber \\ 
      & & \hspace{-0.75cm} \frac{1}{\Delta t \, L^2} \int_{-\frac{1}{2}L}^{\frac{1}{2}L} \! \int_{-\frac{1}{2}L}^{\frac{1}{2}L} \! \int_{t_0 + iT + \Delta t_{\rm f}}^{t_0 + (i+1)T - \Delta t_{\rm l}} \hspace{-0.5cm} f(x,y,z,t) dx\,dy\,dt\;,
   \end{eqnarray}
   where $L = L_x = L_y$, and $\Delta t = T - \Delta t_{\rm f} -
   \Delta t_{\rm l}$ is the length of the subinterval in time. Thus we
   obtain $N$ realisations of the quantity $\overline{f}_i(z)$ that
   can be considered independent. An additional average over these $N$
   values is simply
   \begin{equation}
     \overline{f}(z) = \frac{1}{N} \sum_0^{N-1} \overline{f}_i(z)\;,
   \end{equation}
   where $\overline{f}(z)$ will be the final result. An error estimate
   can be made by computing the standard deviation $\sigma$ of the
   quantity $\overline{f}(z)$ and dividing it by the square root of
   the number of the independent realisations, or to show the full
   range of values attained in the $N$ subintervals. See Fig.
   \ref{pic:variation} for a typical result.


   \section{Results}
   \label{sec:results}

   \subsection{Kinetic helicity}
   Earlier numerical studies of the $\alpha$-effect have shown that
   the FOSA result, Eq.~(\ref{equ:aisofosa}), which states that
   $\alpha$ is proportional to the negative of the kinetic helicity
   holds at least qualitatively (Brandenburg et
   al. \cite{Brandea1990}; Paper I). Recent local helioseimic studies
   have been able to extract information on the kinetic helicity,
   $\mathcal{H}$, in the surface layers of the Sun using the
   correlation of the horizontal components of the divergence and the
   curl of the flow as a proxy (Gizon \& Duvall
   \cite{GizDuv2003}). The result is that for the observed latitude
   range up to the latitude $\Theta \approx \pm 45\degr$, the
   correlation follows a $\cos \theta$ latitude profile. Using
   numerical convection models Egorov et al. (\cite{Egorovea2004})
   showed that the correlation can indeed be used to map the
   helicity. These authors were also able to reproduce the observed
   result with a Coriolis number ${\rm Co} \approx 0.1$ which is
   consistent with the fact that the observations correspond to the
   uppermost 20\,Mm of the solar surface where the Coriolis number is
   of this order of magnitude.

   However, in a recent study K\"apyl\"a et
   al. (\cite{Kaepylaeea2004}) found that the kinetic helicity,
   averaged over the convectively unstable layer, changes sign as the
   Coriolis number is increased to the range expected in the deep
   layers of the solar convection zone, ${\rm Co} \approx 10$. This
   result is in apparent contradiction to analytical theory (R\"udiger
   et al. \cite{Ruedigerea1999}) and the observational and numerical
   results mentioned above. This discrepancy, however, turns out to be
   partly due to the volume averaging performed by K\"apyl\"a et
   al. (\cite{Kaepylaeea2004}), although some apparently new features
   are also present. Fig. \ref{pic:kinhel} shows $\mathcal{H} =
   \overline{\vec{\omega \cdot \vec{u}}}$ for the present calculations
   with Coriolis numbers 1, 4, and 10. It is seen that whereas
   $\mathcal{H}$ is mostly positive at all depths for $\rm{Co = 1}$,
   there is a negative region in the deep layers of the convection
   zone and the overshoot layer for more rapid rotation. The latitude
   trend detected by K\"apyl\"a et al. (\cite{Kaepylaeea2004}) is due
   to the fact that the region of negative helicity expands towards
   the equator whilst the positive peak near the top
   shrinks. Furthermore, the values near the top seem to follow the
   more familiar $\cos \theta$ trend even for ${\rm Co} \approx
   10$. These results suggest that if Eq.~(\ref{equ:aisofosa}) holds,
   the latitudinal behaviour of the $\alpha$-effect should mostly
   change in the deep layers and the overshoot region, but not so much
   in the upper layers of the convection zone. One must bear in mind,
   however, that Eq.~(\ref{equ:aisofosa}) holds only for the isotropic
   case, and that anisotropic expressions derived under FOSA give
   significantly different results and are in better agreement with
   the numerical results (see Sect. \ref{subsec:comFOSA}).

   \begin{figure}
   \centering
   \includegraphics[width=0.5\textwidth]{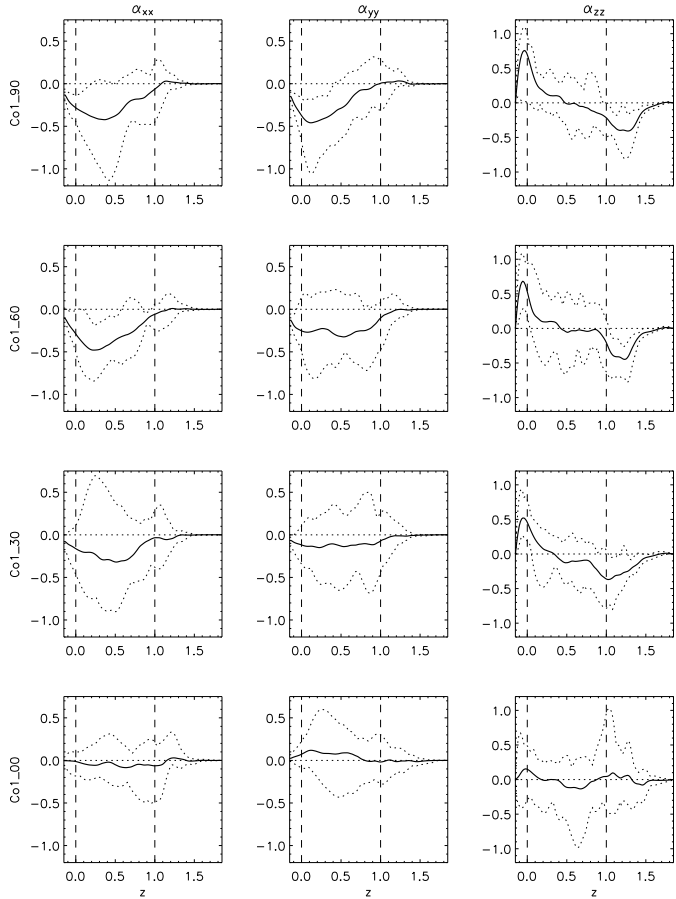}
      \caption{Diagonal components of the $\alpha$-tensor in units of
        $0.01\sqrt{dg}$ for the Co1 set as functions of depth and in
        steps of $\Delta \Theta = 30\degr$ from the south pole (top)
        to the equator (bottom). The dashed vertical lines denote the
        top ($z = 0$) and bottom ($z = 1$) of the convectively
        unstable region. The dotted curves show the full range of
        values attained in the 10 subintervals (see Fig.
        \ref{pic:variation}).}
      \label{pic:alphadiagCo1}
   \end{figure}

   \begin{figure}
   \centering
   \includegraphics[width=0.5\textwidth]{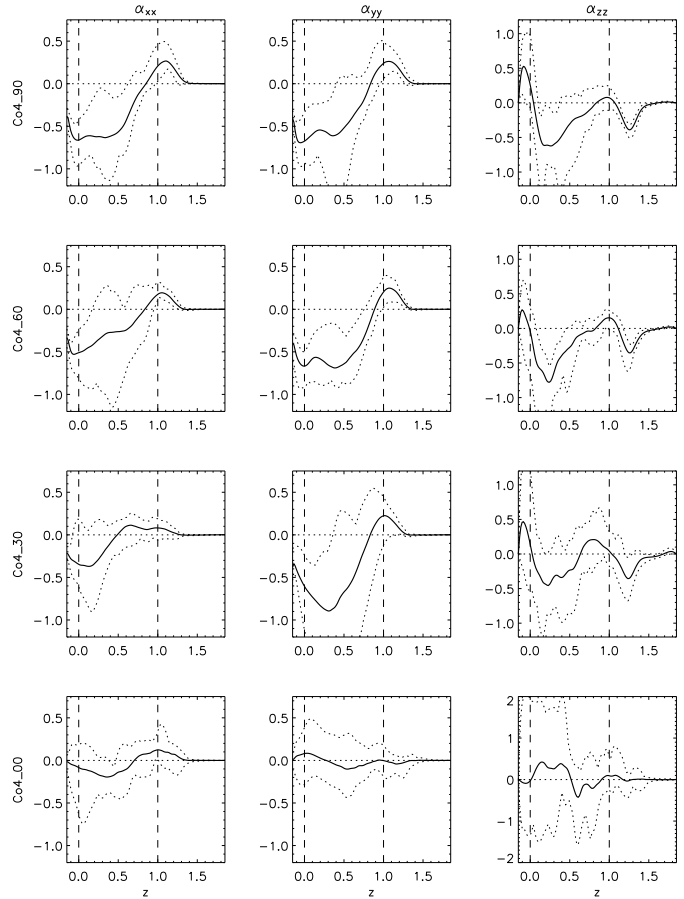}
      \caption{Diagonal components of the $\alpha$-tensor for the Co4
        set, otherwise the same as Fig. \ref{pic:alphadiagCo1}.}
      \label{pic:alphadiagCo4}
   \end{figure}

   \begin{figure}
   \centering
   \includegraphics[width=0.5\textwidth]{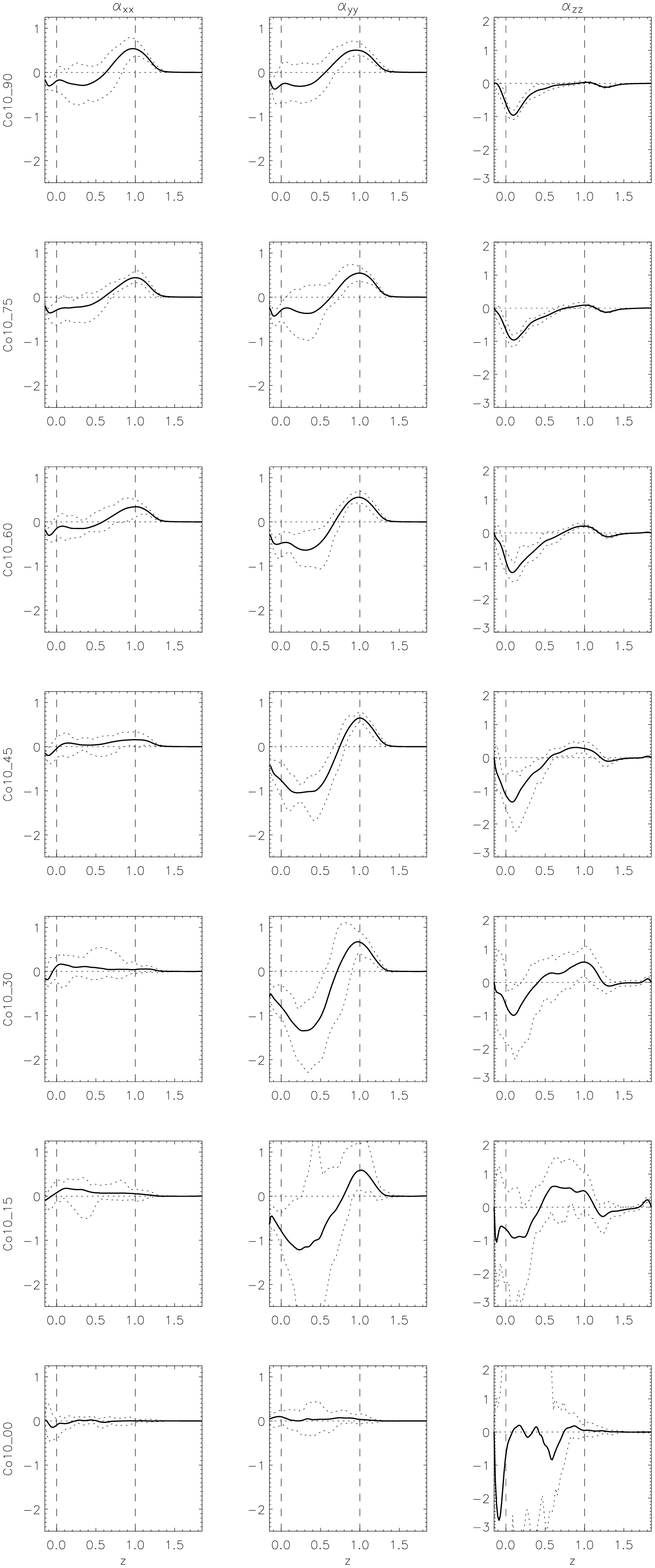}
      \caption{Diagonal components of the $\alpha$-tensor for the Co10
        set in steps of $\Delta \Theta = 15\degr$ from the south pole
        (top) to the equator (bottom). Otherwise the same as
        Fig. \ref{pic:alphadiagCo1}.}
      \label{pic:alphadiagCo10}
   \vspace{-0.5cm}
   \end{figure}

   \subsection{The $\alpha$-effect}
\begin{figure}
   \centering
   \includegraphics[width=0.5\textwidth]{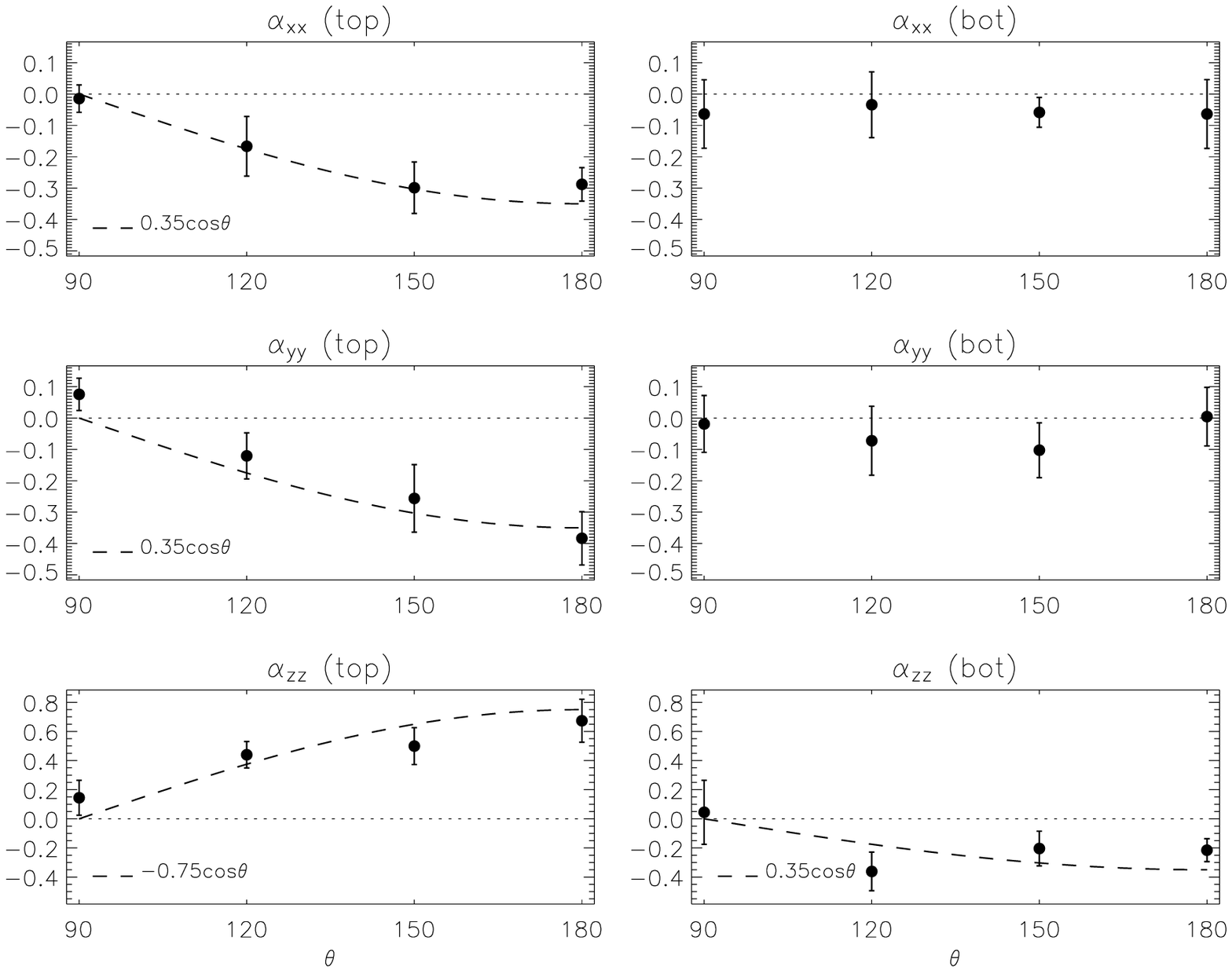}
      \caption{Diagonal components of the $\alpha$-tensor for the Co1
        set as functions of colatitude in units of $0.01\sqrt{dg}$
        from the top (left panels), and bottom (right panels) of the
        convectively unstable region (see Fig.
        \ref{pic:alphadiagCo1}). The error bars denote conservative
        error estimates calculated from the full range
        $\tilde{\sigma}$, denoted by the dotted lines in Fig.
        \ref{pic:alphadiagCo1}, divided by $\sqrt{N}$, where $N =
        10$. The dashed lines show $\cos \theta$-latitude profiles for
        orientation.}
      \label{pic:allatCo1}
   \end{figure}

   The diagonal components of the $a_{ij}$ tensor describe the
   $\alpha$-effect which is responsible for the generation of large
   scale magnetic fields in solar and stellar dynamo models. In other
   contexts, we note that dynamos can also be supported by the
   off-diagonal components of $\vec{\alpha}$ (e.g. R\"adler \&
   Stepanov \cite{RaedlerStepanov2006b}) or via the
   $\overline{\vec{\Omega}} \times \overline{\vec{J}}$ (R\"adler
   \cite{Raedler1969}, \cite{Raedler1986}; Roberts \cite{Roberts1972})
   and $\overline{\vec{W}} \times \overline{\vec{J}}$-effects
   (Rogachevskii \& Kleeorin \cite{RogaKleoo2003},
   \cite{RogaKleoo2004}) that involve rotation and shear flows in
   conjunction with large scale gradients of the mean fields,
   respectively. The present results for $\alpha_{xx}$, $\alpha_{yy}$,
   and $\alpha_{zz}$ correspond to $\alpha_{\theta \theta}$,
   $\alpha_{\phi \phi}$, and $\alpha_{rr}$ in spherical coordinates,
   respectively. Although the present study concentrates on the rapid
   rotation regime, we have also made calculations with smaller
   Coriolis numbers in order to facilitate comparison to earlier
   results. Fig. \ref{pic:alphadiagCo1} shows the $\vec{\alpha}^D$
   components for $\rm Co = 1$. The horizontal components
   $\alpha_{xx}$ and $\alpha_{yy}$ show an expected negative value in
   the convection zone, but basically zero in the
   overshoot. Latitudinal behaviour of the horizontal components is
   consistent with a $\cos \theta$ profile (see Fig.
   \ref{pic:allatCo1}) and is in accordance with the results for the
   kinetic helicity (Fig. \ref{pic:kinhel}).

   This is of interest for the solar dynamo, where the poloidal
   magnetic field is generated from the toroidal one mainly through
   $\alpha_{\phi \phi}$. To lowest order in rotation, this component
   can be represented by $\alpha_{\phi \phi} = \alpha_1 \vec{\hat{g}}
   \cdot \vec{\hat{\Omega}} \propto \cos \theta$, where
   $\vec{\hat{g}}$ and $\vec{\hat{\Omega}}$ are the unit vectors of
   the gradient of turbulence intensity (or equivivalently density),
   and angular velocity, respectively (Krause \& R\"adler
   \cite{KrauRaed1980}). The sign of the vertical component,
   $\alpha_{zz}$, is opposite to those of the horizontal components
   (see also Brandenburg et al. \cite{Brandea1990}; Paper I) and the
   magnitude of this quantity is comparable to the horizontal
   components in the convectively unstable zone and clearly larger,
   and negative, in the overshoot region. The latitude profile is
   consistent with $\cos \theta$ in the convection zone and more or
   less constant in the overshoot layer at least up to $\Theta =
   -30\degr$.

   The results for Co = 4 are very similar to those obtained for Co
   $\approx 2.4$ in Paper II, see Fig. \ref{pic:alphadiagCo4}. Here
   the noteworthy piece of information is the increased anisotropy of
   the $\alpha$s: whereas for Co = 1 the horizontal components were
   both consistent with a $\cos \theta$ latitude profile, for Co = 4
   only $\alpha_{xx}$ follows this trend. $\alpha_{yy}$, on the other
   hand, is basically constant as function of latitude at least up to
   the latitude $-30\degr$. Furthermore, a region of opposite sign
   appears in the overshoot region, consistent with the trend seen in
   the kinetic helicity (Fig. \ref{pic:kinhel}). The vertical
   component $\alpha_{zz}$ predominantly shows a mixed sign, but there
   are indications of a region with negative sign near the top, a
   trend that will emerge much more strongly for more rapid rotation
   (see Fig. \ref{pic:alphadiagCo10}). The latitude trend of
   $\alpha_{zz}$ is similar to that of $\alpha_{yy}$.

\begin{figure}
   \centering
   \includegraphics[width=0.5\textwidth]{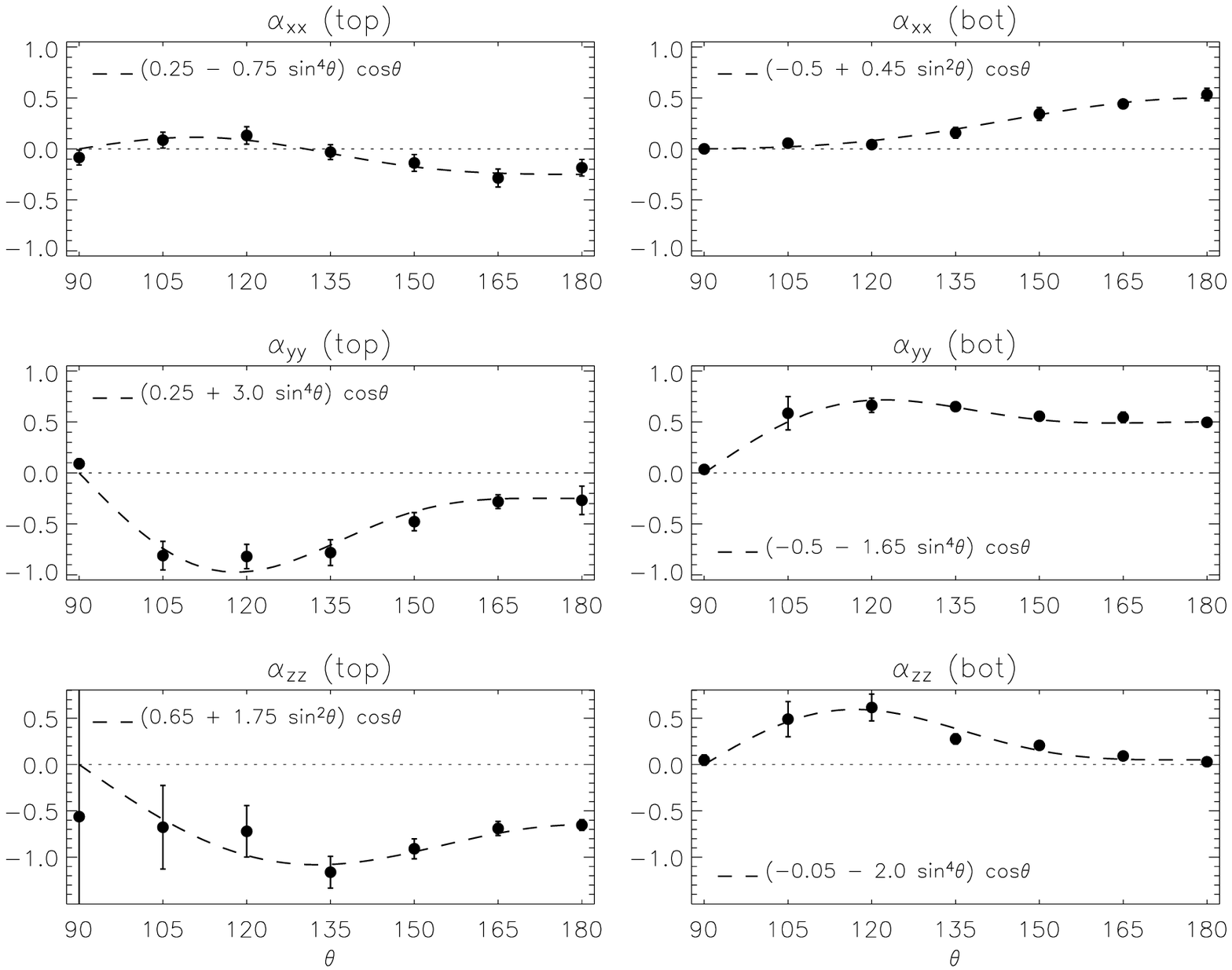}
      \caption{Diagonal components of the $\alpha$-tensor for the Co10
        set as functions of colatitude in units of $0.01\sqrt{dg}$
        from the top (left panels), and bottom (right panels) of the
        convectively unstable region (see Fig.
        \ref{pic:alphadiagCo10}). The dashed curves show the
        expressions denoted in the legends of each panel for
        orientation. Notice the broader plotting range in the panels
        in the lowermost line. The error bars are computed as in Fig.
        \ref{pic:allatCo1}.}
      \label{pic:allatCo10}
   \end{figure}

   The previous numerical studies of the $\alpha$-effect have not gone
   far beyond Co = 1, although, using mixing length models as a guide,
   it can be estimated that the Coriolis number is of the order of 10
   or larger in the deep layers of the solar convection zone
   (K\"apyl\"a et al. \cite{Kaepylaeea2005b}). Fig.
   \ref{pic:alphadiagCo10} presents results for rapid rotation,
   i.e. Co = 10. There are some distinct differences to the models
   with slower rotation. First of all, $\alpha_{xx}$ changes its sign
   between latitudes 45$\degr$ and 60$\degr$ in the convection
   zone. Note that the figure shows the full range of values from the
   ten subintervals, and that the actual error is significantly
   smaller (see Fig. \ref{pic:allatCo10}). The value in the overshoot
   layer, however, still peaks at the pole and diminishes
   monotonically towards the equator, although in a rather steeper
   fashion than $\cos \theta$ (see Fig.  \ref{pic:allatCo10}). It is
   also noteworthy that the magnitude of $\alpha_{xx}$ in the
   convection zone becomes small in comparison to that of
   $\alpha_{yy}$.

   Perhaps the most interesting change that can be observed in the
   rapid rotation regime is the strikingly different latitude
   dependence of $\alpha_{yy}$ (see Fig. \ref{pic:allatCo10}); the
   value in the convection zone now peaks at latitude $-30\degr$,
   whereas the value in the overshoot is basically constant as
   function of latitude all the way down to $\Theta =
   -15\degr$. Furthermore, the sign of the vertical $\alpha$ is now
   totally reversed in comparison to the case of Co = 1. The maximum
   values of $\alpha_{zz}$ in the convection zone and the overshoot
   layer also occur at intermediate latitudes $\Theta = -30\degr
   \ldots -45\degr$.

   We note that the trend of the horizontal $\alpha$-effect as a
   function of rotation at the pole, where $\alpha_{xx} =
   \alpha_{yy}$, is consistent with the behaviour found in Paper
   I. Also the sign change of $\alpha_{zz}$ is reproduced. Analytical
   theory (R\"udiger \& Kichatinov \cite{RuedKic1993}) predicts that
   the horizontal $\alpha$-effect should saturate to a constant value
   when Co is large enough. This seems to occur for $\alpha_{yy}$ in
   the convection zone at and near the pole where the values stay more
   or less constant for all Coriolis numbers. The values in the
   overshoot layer and in the convection zone at lower latitudes,
   however, do not seem to show saturation; according to the study of
   R\"udiger \& Kichatinov (\cite{RuedKic1993}) $\alpha_{yy}$ grows
   roughly $15$ per cent when Co is increased from 4 to 10, whereas in
   the present results the increase is significantly more. It cannot
   be ruled out, however, that signs of saturation could already be
   seen if calculations with intermediate Coriolis numbers were made
   or whether saturation occurs when rotation is more rapid than in
   the present study. The prediction of R\"udiger \& Kichatinov
   (\cite{RuedKic1993}) of a vanishing vertical $\alpha$-effect for
   rapid rotation is not realised in the present results.

   \begin{figure}
   \centering
   \includegraphics[width=0.5\textwidth]{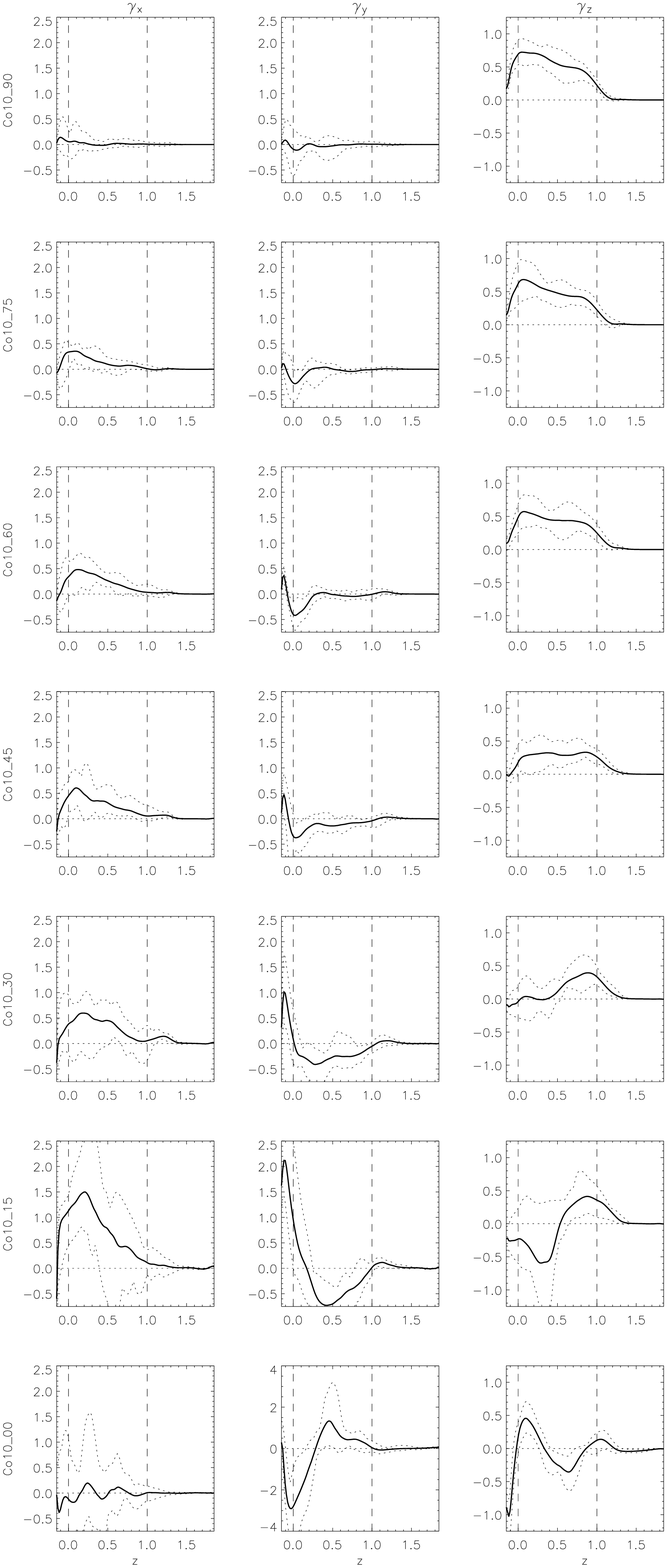}
   \vspace{-0.5cm}
      \caption{The pumping coefficients according to
        Eq.~(\ref{equ:gpump}) from the set Co10 in units of
        $0.01\sqrt{dg}$.}
      \label{pic:gammaCo10}
   \end{figure}

   \begin{figure}
   \centering
   \includegraphics[width=0.5\textwidth]{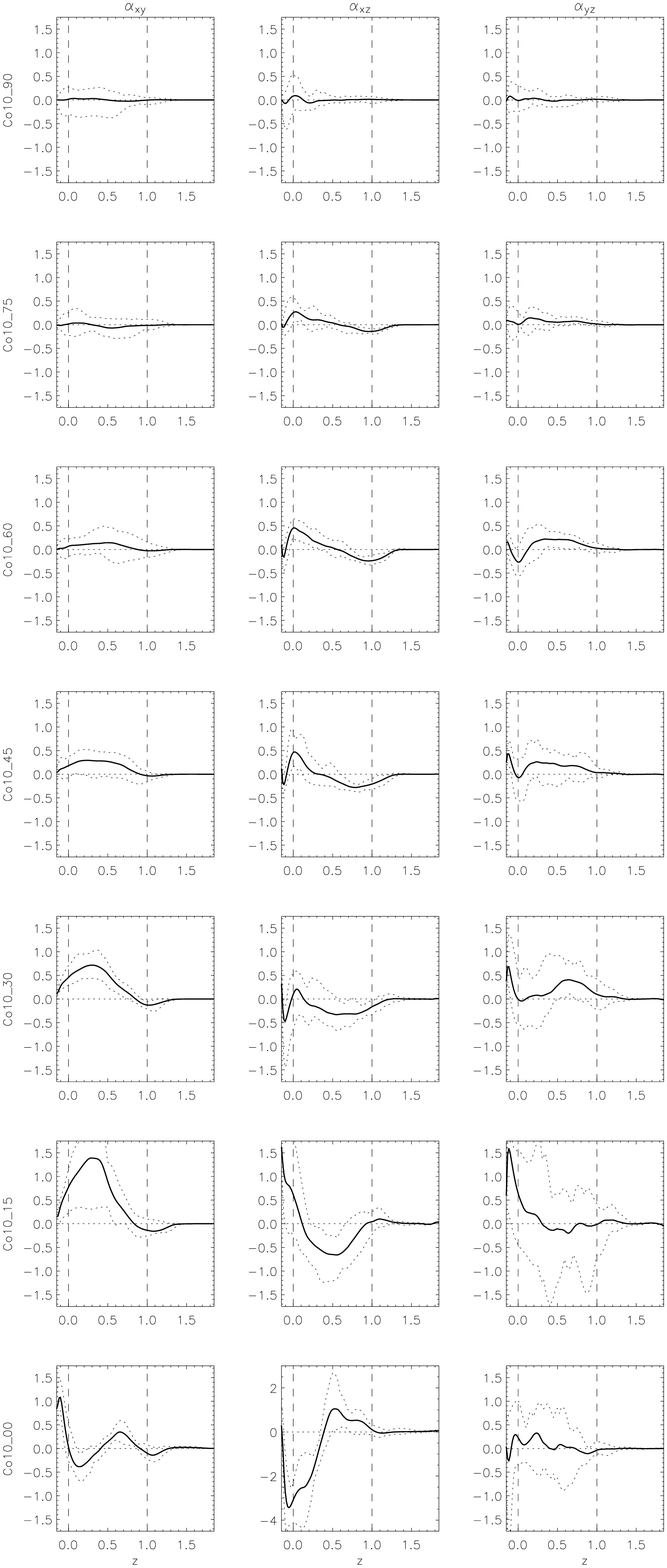}
   \vspace{-0.5cm}
      \caption{Off-diagonal components of the $\alpha$-tensor for the
        Co10 set in units of $0.01\sqrt{dg}$.}
      \label{pic:alphandiagsCo10}
   \end{figure}

   \subsection{Magnetic pumping}
   The existence of field direction dependent pumping of the mean
   field in magnetoconvection calculations was established in Paper
   II. We find that the qualitative results of the present study are
   not significantly different from those obtained in Paper II. The
   main results for the general pumping effect, Eq.~(\ref{equ:gpump}),
   in the rapid rotation case are shown in Fig. \ref{pic:gammaCo10}.
   The latitudinal pumping effect, $\gamma_x$, is postitive,
   i.e. equatorward in the bulk of the convection zone and essentially
   zero in the overshoot layer. The maximum of $\gamma_x$ is near the
   equator, $\Theta = -15\degr$, where the value can be as high as 25
   per cent of the turbulent rms-velocity. If we consider the model to
   describe the deep layers of the solar convection zone where $u
   \approx 10$\,m\,s$^{-1}$ according to mixing length models, the
   latitudinal pumping velocity would correspond to a velocity of
   2.5\,m\,s$^{-1}$, which is of the same order of magnitude as the
   expected values for the equatorward meridional flow in the deep
   layers (e.g. Rempel \cite{Rempel2005}). Taking into account the
   off-diagonal components of the $\alpha$-tensor (Fig.
   \ref{pic:alphandiagsCo10}), it is seen that the equatorward pumping
   of the toroidal field, $\gamma_x^{(y)} = \gamma_{y} + \alpha_{yz}$,
   is further enhanced by the positive $\alpha_{yz}$. The same term
   enters the equation for the latitudinal pumping of the vertical
   field with a negative sign and causes $\gamma_x^{(z)}$ to be
   somewhat smaller, but due to the small magnitude of $\alpha_{yz}$
   in comparison to $\gamma_x$, the qualitative results do not
   change. We note that in comparison to Paper II the equatorward
   trend of the general latitudinal pumping effect $\gamma_x$ is more
   pronounced in the rapid rotation regime.

   The longitudinal component $\gamma_y$ is predominantly retrograde
   in the bulk of the convection zone but often exhibits a thin
   prograde region near the top boundary. The maximum occurs again
   rather near the equator at $\Theta = -15\degr$. The latitude and
   depth distribution of $\alpha_{xz}$ is virtually identical to
   $\gamma_y$ but with different sign. This causes the longitudinal
   pumping of the latitudinal field, $\gamma_{y}^{(x)}$, to become
   very small at virtually all latitudes, whereas the vertical field
   is pumped in the retrograde (prograde) direction in the
   convectively unstable (overshoot) layer with a velocity comparable
   to that of latitudinal pumping velocity.

   The vertical pumping effect is directed mainly downwards, as has
   been found in various earlier numerical studies (Nordlund et
   al. \cite{Nordea1992}; Brandenburg et al. \cite{Brandea1996};
   Tobias et al. \cite{Tobiasea1998}, \cite{Tobiasea2001}; Paper II;
   Ziegler \& R\"udiger \cite{ZiegRued2003}). As was found in Paper II
   the vertical pumping is dominated by the diamagnetic effect, which
   tends to transport the magnetic field from regions of large
   turbulent intensity, characterised by large rms-value of the
   fluctuating velocity, towards regions of lower intensity
   (e.g. R\"adler \cite{Raedler1968}; Krause \& R\"adler
   \cite{KrauRaed1980}). In the anisotropic case (see
   Eqs.~\ref{equ:axyfosa} and \ref{equ:ayxfosa}) the vertical pumping
   is proportional to $-\partial_z \overline{u_z^2}$. There is,
   however, a much stronger latitude dependence of $\gamma_z$ than in
   the more moderate rotation cases studied in Paper II, with even a
   region of upward pumping in the upper layers of the convection zone
   at low latitudes $|\Theta| < 15\degr$. The vertical pumping of the
   latitudinal field, $\gamma_z^{(x)} = \gamma_{x} + \alpha_{xy}$, is
   almost constant as function of latitude in contrast to the strongly
   varying general vertical pumping $\gamma_z$. The pumping of the
   azimuthal field, $\gamma_z^{(y)} = \gamma_{x} - \alpha_{xy}$, on
   the other hand, is directed upward at latitudes $|\Theta| <
   45\degr$.

   \subsection{Comparison to FOSA}
   \label{subsec:comFOSA}
   The exact determination of the transport coefficients, such as
   $a_{ij}$ and $b_{ijk}$, is the most fundamental difficulty in
   dynamo theory. The problem is basically to choose a closure method
   that is both practical and accurate. This problem applies already
   to Eq.~(\ref{equ:mfemf}), where usually only the two first terms
   are taken into account. Some indications already exist for the need
   of higher order derivatives in order to accurately describe the
   electromotive force; see e.g. the numerical geodynamo experiments
   of Schrinner et al. (\cite{Schrinnerea2005}). The second part of
   the closure problem is connected with the derivation of
   $\vec{\mathcal{E}}$ from first principles. The most widely used
   approximation is to consider only the induction equation of the
   small-scale field and neglect all higher than second order
   fluctuations, which is the first-order smoothing approximation (see
   below). It is possible to extend this approach to take into account
   higher order contributions (Nicklaus \& Stix \cite{NickStix1988})
   using the method of a cumulative series expansion (van Kampen
   \cite{vanKampen1974a},\cite{vanKampen1974b}). However, in this
   approach the backreaction of the magnetic field to the flow is
   neglected or at least implicit. Recently, a new approach has been
   introduced (Blackman \& Field \cite{BlackField2002}; see also
   Brandenburg \& Subramanian \cite{BranSub2005a}), where instead of
   $\vec{\mathcal{E}}$ itself, the time derivative
   $\dot{\vec{\mathcal{E}}}$ is computed and the higher order
   contributions, $\vec{T}$, are retained via a damping term $\vec{T}
   = -\vec{\mathcal{E}}/\tau$. This approach is often called the
   `minimal tau-approximation' (hereafter MTA), which leads to
   expressions closely resembling the FOSA results, but with the
   important distinction that due to the fact that the equation of
   motion is also used in the derivation, in addition to the kinetic
   helicity also the small-scale current helicity enters the equation
   for $\alpha$. Recent numerical studies indicate that the current
   helicity is indeed important for the $\alpha$-effect when the
   magnetic field is dynamically important (Brandenburg \& Subramanian
   \cite{BranSub2005b}).

   \begin{figure}[t]
   \centering
   \includegraphics[width=0.5\textwidth]{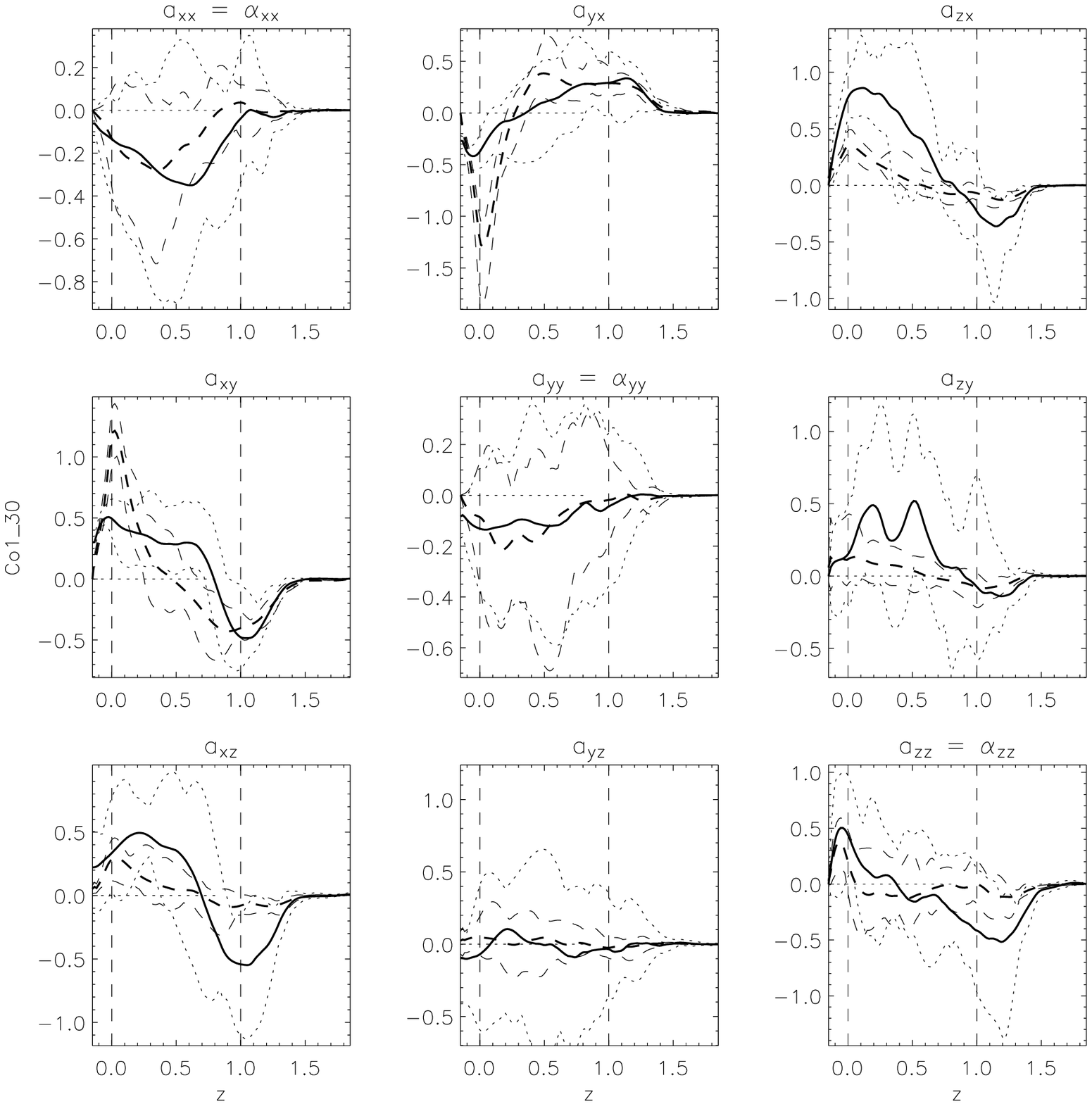}
      \caption{All nine components of the $a_{ij}$-tensor for the run
        Co1-30 (thick solid curves) and the corresponding first order
        smoothing results (thick dashed curves) according to
        Eqs.~(\ref{equ:axxfosa}) to (\ref{equ:azyfosa}). $\tau_{\rm c}
        = 3 \sqrt{d/g}$ in all of the plots. The thin dotted and
        dashed lines give the full range of values for the full
        numerical results and the FOSA equivivalents, respectively.}
      \label{pic:afCo1-30}
   \end{figure}

   \begin{figure}[t]
   \centering
   \includegraphics[width=0.5\textwidth]{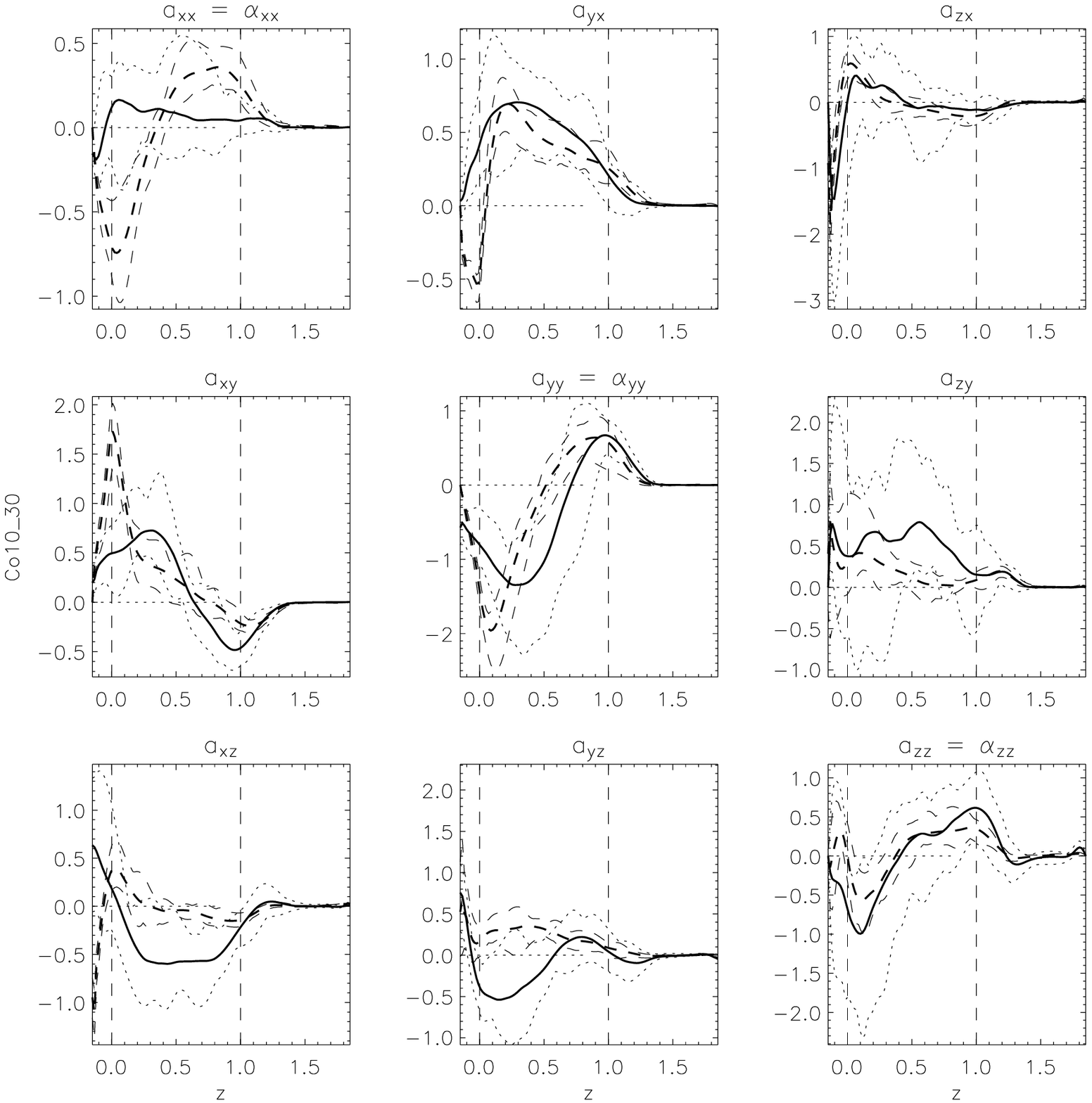}
      \caption{All nine components of the $a_{ij}$-tensor for the run
        Co10-30 (thick solid curves) and the corresponding first order
        smoothing results (thick dashed curves) according to
        Eqs.~(\ref{equ:axxfosa}) to (\ref{equ:azyfosa}). $\tau_{\rm c}
        = 3 \sqrt{d/g}$ in all of the plots. Otherwise the same as
        Fig. \ref{pic:afCo1-30}.}
      \label{pic:afCo10-30}
   \end{figure}

   In order to compare the present numerical results directly to the
   mean-field theory we derive analytical expressions describing the
   coefficients $a_{ij}$ using the first-order smoothing approximation
   (FOSA). Thus we generalise Eq.~(\ref{equ:aisofosa}) to the case of
   anisotropic turbulence. The reason to compare to FOSA is that it is
   relatively straighforward to apply and that the present
   calculations are in the kinematic regime where the influence of the
   magnetic field on the flow is weak and thus the contribution of the
   current helicity is negligible. Furthermore, this comparison will
   be useful in evaluating the applicability of the approximation
   itself in dynamo theory. We start from the equation of the
   fluctuating magnetic field
   \begin{equation}
     \dot{\vec{b}} = \nabla \times (\overline{\vec{U}} \times \vec{b} + \vec{u} \times \overline{\vec{B}} + \vec{G} - \eta \nabla \times \vec{b})\;, \label{equ:flucB}
   \end{equation}
   where
   \begin{equation}
     \vec{G} = \vec{u} \times \vec{b} - \overline{\vec{u} \times \vec{b}}\;.
   \end{equation}
   We drop the term proportional to $\overline{\vec{U}}$ for the time
   being. Furthermore, the diffusion term in Eq.~(\ref{equ:flucB}) can
   be neglected since $\eta \tau_{\rm c}/l_{\rm c}^2 \approx 10^{-3}$,
   where we have used $\tau_{\rm c} = 3 \sqrt{d/g}$ and $l_{\rm c} =
   2\pi/k_{\rm f} \approx d$ (see below). $\vec{G}$ vanishes under the
   assumptions of FOSA, so we are left with
   \begin{equation}
     \dot{b}_i = (\delta_{il}\delta_{jm} - \delta_{im}\delta_{jl}) \partial_j (u_l \overline{B}_m)\;.
   \end{equation}
   The emf is now calculated from 
   \begin{equation}
     \mathcal{E}_i = \varepsilon_{ijk} u_j b_k\;, 
   \end{equation}
   where
   \begin{equation}
     b_k = \int_{-\infty}^{t} \dot{b}_k dt\;. \label{equ:Bk} 
   \end{equation}
   Assuming that the correlation time of the turbulence is short and
   that the mean magnetic field $\overline{\vec{B}}$ varies slowly,
   Eq.~(\ref{equ:Bk}) can be integrated. Following this procedure the
   $a_{ij}$ tensor components can be derived. Using the assumption
   that $\int_0^\infty \overline{u_k(t) \partial_l u_m(t+t')} dt' =
   \overline{u_k(t) \partial_l u_m(t)} \tau_{\rm c}$, with $\tau_{\rm
     c}$ having no dependence on $k$, $l$ or $m$, we find
   \begin{eqnarray}
     a_{xx} & = & \alpha_{xx} = - \tau_{\rm c} (\overline{u_z \partial_x u_y} - \overline{u_y \partial_x u_z})\; \label{equ:axxfosa} \\
     a_{yy} & = & \alpha_{yy} = - \tau_{\rm c} (\overline{u_x \partial_y u_z} - \overline{u_z \partial_y u_x})\; \label{equ:ayyfosa} \\
     a_{zz} & = & \alpha_{zz} = - \tau_{\rm c} (\overline{u_y \partial_z u_x} - \overline{u_x \partial_z u_y})\; \label{equ:azzfosa} \\
     a_{xy} & = &  \tau_{\rm c} (\overline{u_y \partial_y u_z} + \overline{u_z \partial_x u_x}  + \overline{u_z \partial_z u_z})\; \label{equ:axyfosa} \\
     a_{yx} & = & -\tau_{\rm c} (\overline{u_z \partial_y u_y} + \overline{u_x \partial_x u_z}  + \overline{u_z \partial_z u_z})\; \label{equ:ayxfosa} \\
     a_{xz} & = & -\tau_{\rm c} (\overline{u_y \partial_x u_x} + \overline{u_z \partial_z u_y})\; \label{equ:axzfosa} \\
     a_{zx} & = &  \tau_{\rm c} (\overline{u_x \partial_x u_y} + \overline{u_y \partial_z u_z})\; \label{equ:azxfosa} \\
     a_{yz} & = &  \tau_{\rm c} (\overline{u_z \partial_z u_x} + \overline{u_x \partial_y u_y})\; \label{equ:ayzfosa} \\
     a_{zy} & = &  -\tau_{\rm c} (\overline{u_x \partial_z u_z} + \overline{u_y \partial_y u_x})\; \label{equ:azyfosa}
   \end{eqnarray}
   where $\tau_{\rm c}$ is the correlation time. In
   Eqs.~(\ref{equ:axzfosa}) to (\ref{equ:azyfosa}), terms of the form
   $\overline{u_k \partial_x u_k}$ and $\overline{u_k \partial_y u_k}$
   have vanished due to the horizontal periodicity. In practice, we
   use $\tau_{\rm c}$ as a free parameter when comparing with the
   numerical data.

\begin{figure}
   \centering
   \includegraphics[width=0.5\textwidth]{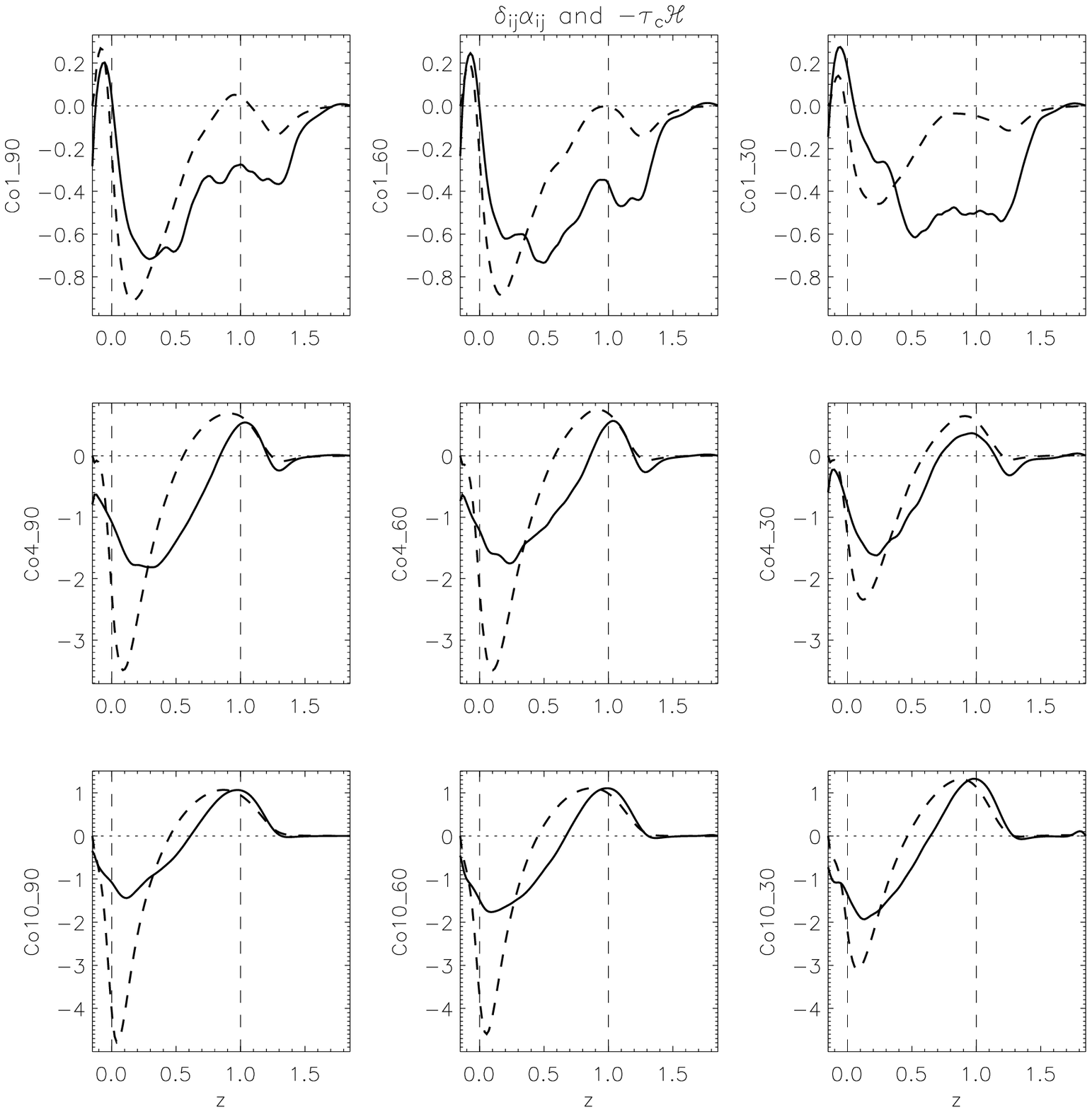}
      \caption{The sum of the diagonal components of $a_{ij}$ (solid
        curves), and $-\tau_{\rm c}\overline{\vec{\omega} \cdot
          \vec{u}}$ (dashed curves) as functions of depth. Both
        quantities are given in units of $0.01\sqrt{dq}$. The leftmost
        column gives data at latitude $-90\degr$, the middle from
        $-60\degr$, and the rightmost from $-30\degr$. The results at
        the equator are not shown. The uppermost panel gives results
        for slow (Co = 1), the middle for moderate (Co = 4), and the
        lower for rapid (Co = 10) rotation.}
      \label{pic:trachel}
   \end{figure}

   Figs. \ref{pic:afCo1-30} and \ref{pic:afCo10-30} show all nine
   components of the $a_{ij}$ tensor for the calculations Co1-30 and
   Co10-30. We use $\tau_{\rm c} = 3 \sqrt{d/g}$ in all of the plots
   and neglect the possible $z$-dependency of $\tau_{\rm c}$. The FOSA
   expressions for the diagonal components of $a_{ij}$ seem to best
   fit to the full numerical results of the corresponding quantities,
   although the fits seem to become poorer when rotation is
   increased. The off-diagonal components $a_{xy}$ and $a_{yx}$ are
   reproduced best in sign and magnitude, whereas the sign, but not
   usually the magnitude, of $a_{xz}$, $a_{zx}$, and $a_{zy}$ is in
   most cases correct. For $a_{yz}$ the correspondence is basically
   non-existent or in the few succesful cases it can be considered
   coincidental. The magnitudes of the off-diagonal components already
   show that a universal correlation time that would fit all the data
   cannot be assigned. We note here that the trace of $a_{ij}$ is
   proportional to the the negative of the kinetic helicity, i.e.
   \begin{eqnarray}
     \delta_{ij} a_{ij} = - \tau_{\rm c} \overline{\vec{\omega} \cdot \vec{u}}\;.
   \end{eqnarray}
   This relation is compared with numerical results in
   Fig. \ref{pic:trachel}. The correspondence is rather good so that
   at least the sign is mostly correct. However, the large peak of the
   helicity near the surface of the convectively unstable region in
   the moderately and rapidly rotating cases is not well reproduced by
   the trace of $a_{ij}$. This discrepancy, however, seems to diminish
   at lower latitudes (see the rightmost panels of
   Fig. \ref{pic:trachel}).

   Although the numerical results differ from the FOSA expressions in
   details, the correspondence is still remarkably good. Very similar
   results were found also for higher Reynolds numbers, with $\rm{Rm}
   \approx 400$ and 580 from preliminary higher resolution
   calculations. It is surprising that the same correlation time gives
   a reasonably good fit for the $\alpha$-effect for all Coriolis
   numbers, although increasing Co is accompanied by a significantly
   decreased correlation time of the convective motions (see
   K\"apyl\"a et al. \cite{Kaepylaeea2006}) and a similar decrease is
   also seen when MTA is applied to the Reynolds stresses (K\"apyl\"a
   et al. \cite{Kaepylaeea2005a}). We can also estimate the Strouhal
   number from
   \begin{equation}
     {\rm St} = u_{\rm rms} k_{\rm f} \tau_{\rm c}\;,
   \end{equation}
   where $k_{\rm f}$ is the wavenumber where kinetic energy spectrums
   peaks. The exact value of $k_{\rm f}$ is difficult to assess
   accurately since it tends to vary in time and also in
   depth. However, we note that the power always peaks in rather large
   scales\footnote{The smallest possible horizontal wavenumber that
     fits in the box is $k = 2\pi/L_x = \pi/2$.}, i.e. $k_{\rm f} = (5
   \ldots 10)\,d^{-1}$, with which we obtain values of St which are of
   the order of unity or larger. As in K\"apyl\"a et
   al. (\cite{Kaepylaeea2006}) we find that $u_{\rm rms}$ decreases
   and $k_{\rm f}$ increases as function of rotation, effectively
   balancing each other so that St is more or less constant as
   function of Co. The values of the Strouhal number obtained by using
   the correlation time from the fits to analytic FOSA expressions
   therefore suggest that higher than second order terms should be
   important for the dynamo coefficients already in the kinematic
   regime.


\section{Conclusions}
\label{sec:conclu}
   We use three-dimensional modelling of convection in rectangular
   boxes at different latitudes and Coriolis numbers to study the
   $\alpha$-effect and turbulent magnetic pumping. The results for the
   $\alpha$-effect at moderate and intermediate rotation can be
   summarised as follows
   \begin{itemize}
   \item As in previous studies (Papers I and II) we find that the
     horizontal components $\alpha_{xx}$ and $\alpha_{yy}$ are
     negative in the convection zone on the southern hemisphere and
     follow a more or less $\cos \theta$ latitude profile for moderate
     rotation (${\rm Co} = 1$). The latitude trend is in accordance
     with the similar distribution of the kinetic helicity, which, in
     the simpler case of isotropic turbulence is proportional to the
     negative of the (scalar) $\alpha$-effect. Values of the
     horizontal $\alpha$s in the overshoot region are consistent with
     zero.
   \item
     For Co = 4 we find that the latitude trend for $\alpha_{xx}$
     remains effectively the same as for Co = 1, whereas the magnitude
     of the quantity increases. For the same Coriolis number, the
     component $\alpha_{yy}$ is more or less constant as function of
     latitude at least up to $\Theta \approx -30\degr$. A pronounced
     region of oppositely signed horizontal $\alpha$s appear in the
     overshoot region with latitude distribution similar to the
     corresponding quantity in the convection zone.
   \item The vertical component $\alpha_{zz}$ is positive in the
     convection zone and negative in the overshoot for Co = 1 and
     exhibits more sign changes as a function of depth for Co = 4.
   \end{itemize}
   As noted recently by K\"apyl\"a et al. (\cite{Kaepylaeea2004}), the
   qualitative behaviour of kinetic helicity changes significantly
   when rotation is rapid enough, i.e. ${\rm Co} \approx 10$, raising
   the question of the $\alpha$-effect in this regime. This regime
   also coincides with the deep layers of the solar convection
   zone. New results for the rapid rotation regime include (see also
   Fig. \ref{pic:allatCo10}).
   \begin{itemize}
   \item The $\alpha_{xx}$ component shows diminishing magnitude and a
     sign change in the convection zone near the equator for Co = 10
     and a steeper than $\cos \theta$ trend in the overshoot.
   \item In the stellar context the most interesting component is
     $\alpha_{yy}$, corresponding to $\alpha_{\phi \phi}$ in spherical
     coordinates. Unlike the earlier studies where rotation is not so
     strong, $\alpha_{yy}$ in the convection zone no longer peaks at
     the poles, but rather much closer to the equator at $\Theta =
     -30\degr$. The value in the overshoot is virtually constant as
     function of latitude.
   \item Another new feature is the sign change of $\alpha_{zz}$,
     which now coincides with the azimuthal component. Also the
     latitudinal distribution is similar, convection zone and
     overshoot values peaking around $\Theta = -30\degr \ldots
     -45\degr$.
   \item The maximum values of the diagonal components of $a_{ij}$ are
     of the order of $0.2-0.3\,u_{\rm rms}$. Considering the models to
     describe the deep layers of the solar convection zone where $u
     \approx 10$\,m\,s$^{-1}$, the magnitude of these terms would be
     of the order of $2-3$\,m\,s$^{-1}$.
   \end{itemize}
   The turbulent pumping effects can be summarised as follows
   \begin{itemize}
   \item The latitudinal pumping is equatorward and peaks near the
     equator, i.e. around $\Theta = -15\degr$. The magnitude of
     $\gamma_x$ is maximally about 0.2\,$u_{\rm rms}$, which would
     corresponds to $2-3$\,m\,s$^{-1}$ in the deep layers of the solar
     convection zone.
   \item Longitudinal pumping is retrograde in the convection zone and
     prograde near the upper boundary of the domain. At the equator
     the prograde region is larger.
   \item Vertical pumping is predominantly directed downwards and
     dominated by the diamagnetic effect. Whereas the poloidal field
     is always pumped downwards, the azimuthal field is pumped upwards
     at low latitudes, $|\Theta| < 45\degr$.
   \end{itemize}

   The present new results are likely to have consequences for
   mean-field models of the solar dynamo. First of all, a
   long-standing problem of solar dynamo models is that the radial
   differential rotation is strongest at high latitudes (e.g. Schou et
   al. \cite{Schouea1998}) and the often assumed simple form of the
   $\alpha$-effect, proportional to $\cos \theta$, imply that the
   dynamo is most efficient at the poles, but virtually no sunspots
   are observed at latitudes higher than $|\Theta| > 40\degr$. The
   tendency of $\alpha_{yy}$ to peak at latitudes around $|\Theta| =
   30\degr$ in the regime where the rotational influence is comparable
   to the bottom of the solar convection zone is likely to alleviate
   this problem. The second major problem for solar dynamo models is
   due to the positive radial gradient of the angular velocity near
   the equator (e.g. Schou et al. \cite{Schouea1998}) which, with a
   positive $\alpha$-effect in the northern hemisphere, leads to
   poleward migration of the activity belts (Parker
   \cite{Parker1987}), in contradiction with observations. The
   latitudinal pumping will most likely also help in alleviating this
   problem as well, and it has been noted that downward pumping has a
   similar effect in mean-field models (Brandenburg et al.
   \cite{Brandea1992}). The longitudinal pumping velocity
   $\gamma_{\phi}$ plays no role in axissymmetric dynamo models,
   although its gradients $\nabla_r \gamma_{\phi}$ and $\nabla_\theta
   \gamma_{\phi}$ give rise to terms analogous to those that appear
   due to differential rotation. The magnitude of this effect,
   however, is likely to be small in comparison to that of
   differential rotation. Longitudinal pumping may be important in
   more general configurations where, for example, the propagation of
   non-axissymmetric structures could be explained by this effect. The
   motivation for this is that in the case of the Sun evidence for
   relatively short-lived nonaxisymmetric structures exists (of the
   order of 10 rotations from sunspot statistics by Pelt et
   al. (\cite{Peltea2006}) up to the length of the sunspot cycle of 11
   years from solar flares by Bai \cite{Bai2003}), whilst for rapidly
   rotating late-type stars the concept of active longitudes
   persisting over several stellar cycles is widely accepted (see
   e.g. Berdyugina \& Tuominen \cite{BerdyTuo1998}).

   In order to investigate the applicability of the first-order
   smoothing approximation (FOSA) we have derived analytical
   expressions of the transport coefficients which we compare to the
   quantities obtained numerically. We find rather good correspondence
   for the $\alpha$-effect in the moderate rotation case (Co = 1),
   whereas the off-diagonal components of $a_{ij}$ are only in
   qualitative accordance or fail to reproduce even that. For more
   rapid rotation the fits are poorer, but still manage to capture the
   qualitative features correctly. The correlation time, used here as
   a free parameter to fit the numerical data to the analytical
   expressions, turns out not to have a universal value. Despite of
   this, we have used $\tau_{\rm c} = 3 \sqrt{d/g}$ for the whole
   parameter range investigated, as it seems to reproduce the
   $\alpha$-effect satisfactorily for all rotation rates
   investigated. With the correlation times obtained from the fitting
   procedure we can also estimate the Strouhal number, for which we
   found values of the order of unity independent of rotation. The
   lack of dependence on rotation results in directly from the
   universal correlation time used in the fitting which, however, is
   not consistent with the correlation time computed directly from the
   flow (K\"apyl\"a et al. \cite{Kaepylaeea2005a},
   \cite{Kaepylaeea2006}). At the moment the reason for this behaviour
   is unclear.

   In the present study the $\alpha$-quenching (e.g. Cattaneo \&
   Vainshtein \cite{CattVain1991}; Vainshtein \& Cattaneo
   \cite{VainCatt1992}) issue has not received any attention, but we
   note that exploration of this issue is further required. A related
   problem concerns the magnetic helicity fluxes (Kleeorin et
   al. \cite{Kleeorinea2000}; Vishniac \& Cho \cite{VishCho2001}) and
   the effects of boundary conditions (Brandenburg \& Sandin
   \cite{BranSandin2004}) on the efficiency of the
   $\alpha$-effect. Furthermore, to date, the $b_{ijk}$-tensor
   components have not to our knowledge been calculated from numerical
   models of convection. These points, however, cannot be addressed
   within the scope of the present study but should be explored in the
   future.

\begin{acknowledgements}
   PJK acknowledges the financial support from the graduate school for
   astronomy and space physics of the Finnish academy and the
   Kiepenheuer-Institut for travel support. Furthermore, the Academy
   of Finland grant 1112020 is acknowledged. The authors wish to thank
   the referee (Prof. K.-H. R\"adler) for his thorough and critical
   reading of the manuscript and his numerous comments which helped to
   clarify and improve the paper significantly.
\end{acknowledgements}

\end{document}